\documentclass[openacc]{rstransa}
\jname{rsta}
\Journal{Phil. Trans. R. Soc}

\newcommand{\He}{$^3$He}

\newcommand{\be}{\begin{equation}}
\newcommand{\ba}{\begin{align}}
\newcommand{\bs}{\begin{split}}
\newcommand{\ee}{\end{equation}}
\newcommand{\ea}{\end{align}}
\newcommand{\es}{\end{split}}
\newcommand{\reffig}[1]{Fig.~\ref{#1}}

\newcommand{\grad}{\mbox{\boldmath$\nabla$}}
\newcommand{\Del}{{\mbox{\footnotesize $\Delta$}}}
\newcommand{\vDelta}{\mbox{\boldmath$\Delta$}}

\newcommand{\hDelta}{\hat{\Delta}}

\newcommand{\hg}{\hat{g}}
\newcommand{\hf}{\hat{f}}
\newcommand{\vsigma}{\mbox{\boldmath$\sigma$}}
\def\vf{{\bf f}}
\def\vj{{\bf j}}
\def\vp{{\bf p}}
\def\vq{{\bf q}}
\def\vQ{{\bf Q}}
\def\vR{{\bf R}}
\def\vv{{\bf v}}
\def\cY{{\cal Y}}
\def\hp{{\hat{p}}}
\def\hvp{{\hat{\bf p}}}
\def\ket#1{\mbox{$\displaystyle\vert\,#1\,\rangle$}}
\def\pder#1#2{\mbox{$\displaystyle\frac{\partial #1}{\partial #2}$}} 
\def\figref#1{Fig.~\ref{#1}}

\begin{document}

\title{Andreev Bound States in Superconducting Films and Confined Superfluid \He}

\author{
Anton~B.~Vorontsov$^{1}$
}

\address{
$^{1}$Department of Physics, Montana State University, Bozeman, Montana 59717, USA 
}

\subject{physics, condensed matter, superconductivity}


\corres{Anton Vorontsov\\
\email{avorontsov@physics.montana.edu}}

\begin{abstract}
This paper reviews the confinement-driven phase transitions, and the 
prediction of superfluid phases with broken time-reversal 
or translational symmetry in thin films. 
The new phases are a result of particle-hole coherent Andreev scattering processes 
that create 
quasiparticle states with energies inside the superconducting gap. 
These states cause profound restructuring of the low-energy spectrum in the 
surface region of several coherence lengths $\xi_0$ with large 
spatial variations of the superconducting order parameter. 
In confined geometry, such as slabs, films, pores, or nano-dots, 
with one or more physical dimensions $D \sim 10 \xi_0$, 
Andreev bound states can dominate properties of the superfluid phases, 
leading to modified 
experimental signatures. 
They can dramatically change the energy landscape, and 
drive transitions into new superfluid phases, that are typically unstable in the bulk 
superfluid. 
On the examples of single-component singlet $d$-wave superconductor, 
and triplet multi-component superfluid \He\ I show how properties 
of condensed phases in restricted geometry depend on the order parameter structure. 
I will highlight the connection between Andreev bound states and 
confinement-stabilized phases with additional broken symmetries, 
describe recent progress and open questions in theoretical 
and experimental investigation of superfluids in confined geometry. 
\end{abstract}


\begin{fmtext}
\end{fmtext}

\maketitle



\section{Introduction}
\label{sec:intro}

Majority of known superconductors, 
including ones with the highest and lowest measured transition temperatures, 
as well as superfluid \He, belong to the class of unconventional pairing states.\cite{Sigrist:1991ua}
By definition, 
they break multiple symmetries of the 
high-temperature normal phase, that usually include 
the $U(1)$ symmetry, rotations and reflections 
determined by the crystal symmetry group, 
and rotations in spin space. 
These superconductors develop order parameter (OP) that beside having a definite phase resulting from 
broken global $U(1)$ symmetry, possess non-trivial structure in momentum and spin spaces. 
These broken symmetries manifest themselves in characteristic 
low-energy spectrum of quasiparticles, 
and show up in thermodynamic, magnetic, transport properties of 
unconventional materials. 

The low-energy excitations above the ground state of a condensate of Cooper pairs, 
are combinations of particles and holes, 
that can be written as a sum of creation and annihilation operators,
$
\alpha^\dag_{\vp s} \ket{G.S.} = 
\left( u^*_{\vp\; ss'} a_{\vp s'}^\dag - 
 v^*_{\vp\; ss'} a_{-\vp s'} \right) \ket{G.S.} = E_{\vp} \ket{G.S.}
$,
with excitation energies
$E_{\vp} = \sqrt{v_f^2(p-p_f)^2 + |\Delta(\vp_f)|^2}$ 
indirectly reflecting 
broken symmetries of the 
order parameter $\Delta(\vp_f)$ and type of the pairing interactions. 

Scattering of these quasiparticles on the boundaries, surfaces or interfaces,
lead to further modification of the condensate properties. 
Spatial variations of the superconducting 
order parameter $\Delta(\vR,\vp_f)$,
caused by the boundary conditions or other competing effects, result in a new kind of scattering processes. 
The quasiparticles 
can convert between particles and holes, transferring weight between (spin-matrix) amplitudes 
$u_{\vp\; ss'} \leftrightarrow v_{\vp \; ss'}$ \cite{Andreev:1964wk}.
These scattering processes originate in the particle-hole coherence of superconductors,
and they play important role in non-uniform superconducting environments. 
These are the most important 
scattering events for the low-energy sector, 
resulting in new quasiparticle states 
below the bulk gap edge and  
bound to the inhomogeneous regions, 
whose properties will dominate the 
physics of the boundary region.\cite{Herath:1990gd,KIESELMANN:1983uq,ZHANG:1988tb} 
The new 
quasiparticle spectrum leads to 
strong order parameter variations on 
the coherence length scale $\xi_0 =\hbar v_f / 2\pi k_B T_c$. 

The new quasiparticle states that arise in the boundary regions can be broadly divided into 
two categories. 
One type of states is determined by the properties of the superconductors 
in the bulk, and largely independent of the order parameter suppression near the surface. 
In that sense they are determined by the far away regions, as in a domain wall 
and thus have a more of a topological character, similar to mass term change in 
relativistic Dirac equation \cite{JACKIW:1976tv}, 
These states can have any energy-momentum dispersion, depending on the 
orientation of the surface, and initial and final values of the order 
parameter along incoming and outgoing quasiparticle's trajectories. 
The second class is the one that was considered by Andreev originally, 
where quasiparticles experience multiple reflections inside the effective potential well 
created by the suppression of the OP amplitude. 
This typically leads to bound states with energies not too far below the 
gap edge. 

Quasiparticle states with subgap energies, bound to the surface region 
of several coherence lengths,
have a profound effect on physical properties.
Examples include presence of a subdominant channel $d+is$ near surface that 
leads to splitting of the 
zero-energy states and generation of current-carrying state 
that breaks time-reversal symmetry.\cite{MATSUMOTO:1995tf,Fogelstrom:1997wp}
Surface ABS carry paramagnetic currents that run in the opposite direction to the 
usual superconducting screening currents, which shows up in the 
anomalous behavior of the penetration depth.\cite{Fogelstrom:1997wp,Walter:1998uw} 
Interaction of bound states with self-induced magnetic field leads to lowering of 
energy in the surface region and spontaneous 
generation of currents on penetration length scale.
\cite{Higashitani:1997bv,Kusama:1999wp,Barash:2000vt,Lofwander:2000vh}
In quantum wires, Andreev bound states (ABS) can lead to new pairing states with 
spin-triplet character.\cite{Bobkov:2004iz}

Even more profound are effects of the bound states on the properties 
of superconductors and superfluids in confined geometry, such as 
films, slabs, pores and nano-dots. If geometrical dimensions of a 
sample are several coherence lengths, the order parameter suppression 
is significant in the entire volume, and the spectrum of low-energy 
excitations is very different from that of bulk superfluid, 
is mostly dominated by the quasiparticle 
states that scatter off the surfaces. 
This results in a significant modifications of the thermodynamic and 
non-equilibrium properties, and shows up in NMR, heat capacity, collective mode dynamics, 
and other experimental probes. 

Moreover, due to significantly modified order parameter and 
quasiparticle spectrum, the landscape of free energy is changed as well. 
Additional constraints imposed by the boundary conditions 
make bulk phases considerably less favorable, and make room for new 
phases that have the lowest energy state in the new landscape. 
One of the most interesting possibilities is appearance of states that 
have symmetry or topological properties different from those of the 
bulk phases. 

The many effects associated with the bound states, and their sensitivity to 
the nature of the pairing, surface geometry, makes investigation of 
properties of superfluids in confinement both challenging and interesting. 
Spatially constrained superfluids and geometry manipulation
provide new ways to learn about properties of 
unconventional superconductors, insight into the nature of superconducting pairing, 
and possibility of generating new superfluid phases. 
Manipulation of the surface states can change electronic transport across 
superconductor-normal interfaces and promise better control of small-scale superconducting systems, 
opening possibilities to 
utilize them in new devices. 

In this review I describe properties of unconventional superconductors 
in thin films, and highlight the connection between 
unusual properties of the confined superfluids with the presence and 
structure of Andreev bound states spectrum. 
In section \ref{sec:theory} a brief summary of theoretical approach to 
study non-uniform superconductors is given. 
In section \ref{sec:new} 
I discuss the structure of new phases that are expected to 
appear in thin films of unconventional superconductors and 
superfluid \He. 
Finally, in part \ref{sec:experim}, 
I outline recent development in experimental techniques that 
are oriented to better understanding and control of 
quasiparticle states and superfluid phases in confinement.


\section{Surface bound states near surfaces and in domain walls}
\label{sec:theory}

Investigation of superfluid condensates in confined geometry requires 
careful treatment of multiple aspects of the physics, that influence the 
energy balance.
To calculate details of the order parameter suppression precisely,  
one needs to specify scattering properties of 
the surfaces, their shape and orientation, and size of the container. 
Other parameters, such as dimensionality of the geometry, shape of the 
Fermi surface, external fields or other pairbreaking effects, 
and strong-coupling corrections, all can affect the energetics and may favor different phases. 

Early theoretical investigation of pairing states in 
constrained geometry focused on properties of 
superfluid \He. 
Boundary conditions for the order parameter in the A-phase were proposed in \cite{AMBEGAOKAR:1974wl}, 
and suppression of the transition temperature calculated \cite{KJALDMAN:1978vo}, 
based on de Gennes' formulation of inhomogeneous superfluidity
in terms of semiclassical correlation functions. 
Later investigation of superfluid phases in slabs and cylindrical pores
used Ginzburg-Landau (GL) approach \cite{Li:1988tu,Fetter:1988bl,Wiman:2013ky,Aoyama:2014uc}.
Properties of the superflow and NMR responses of confined \He\ were 
calculated \cite{Ullah:1988uw}. 
Although the GL equations have a limited applicability range, 
they have an advantage of being the simplest approach to inhomogeneous 
problems in the long-wavelength limit, and can easily include the strong-coupling corrections 
via phenomenological parameters. 

A more sophisticated technique to address strongly non-uniform states is based 
on the quasiclassical Green's functions.\cite{Eilenberger:1968wb,Larkin:1969wa,Serene:1983vc}
The quasiclassical theory has been used to study confined superfluids extensively, since 
it is applicable to various phenomena under a broad range of conditions, including
arbitrary temperatures, fields, and systems out of equilibrium. 
\He\ flow and superfluid density in film geometry have been investigated in 
\cite{Kopnin:1986cv,Buchholtz:1993hr,Buchholtz:2000bt,Yamamoto:2000wp}; 
the A-B transition in thin films was discussed in 
\cite{Hara:1988di,Nagato:2000vv,VorontsovAB:2003ki};
a detailed analysis of thermodynamic properties of the 
A-phase was presented in \cite{VorontsovAB:2003ki}; 
study of thermodynamic properties and Majorana signatures 
of distorted \He-A and \He-B phases 
in narrow channels and slabs was done in \cite{Tsutsumi:2011eo,Wu:2013ev}.

For completeness we briefly summarize the main points of the technique: 
the quasiclassical propagator (Green's function) 
$\widehat{g}(\vR,\vp_f; \epsilon)$ describes 
quasiparticle correlations on the scale  
$\hbar/p_f \ll |\delta \vR| \sim \xi_0= \hbar v_f/2\pi k_B T_c$, with energy $\epsilon$, 
along a quasiparticle's classical trajectory defined by velocity $\vv_f$ on the Fermi surface 
at point $\vp_f$. It satisfies Eilenberger transport equation and normalization condition:
\begin{align}\label{eq:eilenberger}
\begin{split}
[\epsilon \widehat\tau_3 - \widehat\Delta , \widehat{g}] + i\hbar\vv_f\cdot\grad_{\vR} \widehat{g} = 0
\quad,\qquad
\widehat{g}^2 = -\pi^2 \,.
\end{split}
\end{align}
The propagator has $4\times4$ matrix structure in particle-hole and spin space, denoted here by wide hat.
We only consider mean-field self-energies that describe the superconducting order parameter, 
\begin{equation}
\widehat{g} = 
\left( \begin{array}{cc} \hg & \hf \\ \underline{\hf} & \underline{\hg} \end{array} \right)\,,
\qquad\qquad
\widehat\Delta = \left(\begin{array}{cc}
0 & \hat\Delta_\vp  \\
{\underline{\hat\Delta}}_\vp & 0 
\end{array}\right) \,,
\end{equation}
where the $2 \times 2$ spin structure of the order parameter, denoted by narrow hat, 
for singlet state is 
$ \hat\Delta = \Delta_{\vp} (i\sigma_y)$ 
and for triplet is 
$ \hat\Delta = \vDelta_{\vp} (i\vsigma\sigma_y)$ 
The energy can be shifted up or down in the complex plane to obtain 
Retarded ($\epsilon+i0$),  
Advanced ($\epsilon-i0$), 
or finite temperature Matsubara Green's functions 
($\epsilon \to i\epsilon_m = i \pi T(2m+1))$. 
The particle-hole components of the propagator are related through symmetry \cite{Serene:1983vc}
that we denote by underline-operation, which we write for complex energy 
that combines both Retarded/Advanced and Matsubara representations: 
\be
\underline{\hat\alpha}(\vR,\vp_f; \epsilon + i\epsilon_m) = 
\hat\alpha(\vR,-\vp_f; -\epsilon + i\epsilon_m)^* \,.
\label{eq:tau1symm}
\ee
This symmetry relates objects in the same half-plane of the complex energy.  
In addition, there is another symmetry that relates 
propagator components in the upper and lower 
half-planes:
\be
\hat{g}(\vR,\vp_f; \epsilon + i\epsilon_m)^\dag  = 
\hat{g}(\vR,\vp_f; \epsilon - i\epsilon_m) 
\;,\qquad
\hat{f}(\vR,\vp_f; \epsilon + i\epsilon_m)^\dag  = 
-{\underline {\hat f}}(\vR,\vp_f; \epsilon - i\epsilon_m)  \,.
\label{eq:tau3symm}
\ee


Calculation of the equilibrium order parameter is most conveniently done using Matsubara technique. 
Equation (\ref{eq:eilenberger}) for $\widehat g$ has 
to be solved self-consistently with the equation for the order parameter. 
If the pairing interaction is separable with basis functions 
in momentum space $\cY(\vp_f)$, this equation for singlet pairing has the form:
\be
\Delta(\vR,\vp_f) = 
T \sum_{|\epsilon_m| < \Lambda} 
N_f\; \left< 
 V \; \cY(\vp_f) \cY^*(\vp_f') \; f(\vR,\vp_f'; i\epsilon_m)
\right>_{\vp_f'}
\ee
with attractive pairing interaction $V>0$, cut off at energy $\Lambda$. 
Angle brackets traditionally denote Fermi surface integration, and $N_f$ is density of states 
at the Fermi level per one spin projection. 

The most convenient numerical route to solve the transport equations 
(\ref{eq:eilenberger}) is 
to use parametrization of the Green's function in terms of the 
coherence amplitudes, that are chosen to automatically satisfy the 
normalization condition. 
Following notation in \cite{Eschrig:2000ux}: 
\begin{equation}
\widehat{g}(\vR,\vp_f; \epsilon) = \mp i \pi 
\left(\begin{array}{cc}
(1 - \hat{\gamma} \hat{\underline\gamma})^{-1} & 0 \\
0 & (1 - \hat{\underline\gamma} \hat{\gamma})^{-1}
\end{array}\right)
\left(\begin{array}{cc}
(1 + \hat\gamma \hat{\underline\gamma})  & 2 \hat\gamma \\
-2\hat{\underline\gamma} & -(1 + \hat{\underline\gamma} \hat{\gamma})
\end{array}\right)
\,,
\end{equation}
where $\epsilon=\epsilon' + i\epsilon''$ 
is fully complex, and $(-1)$ sign applies to upper half plane $\epsilon'' > 0$, 
while $(+1)$ sign is used for $\epsilon'' < 0$ functions. 
The coherence amplitudes are $2 \times 2$ matrices in spin space, satisfying symmetries 
that follow from (\ref{eq:tau1symm}) and (\ref{eq:tau3symm})
\begin{align}
\begin{split}
&\underline{\hat\gamma}(\vR,\vp_f; \epsilon' + i\epsilon'') = 
\hat\gamma(\vR,-\vp_f; -\epsilon' + i\epsilon'')^* 
\,,\\ 
&{\underline {\hat \gamma}}(\vR,\vp_f; \epsilon' + i\epsilon'')  =
\hat{\gamma}(\vR,\vp_f; \epsilon' - i\epsilon'')^\dag  
\,,
\end{split}
\label{eq:gammasymm}
\end{align}
and obeying non-linear 
differential equation of Riccati type, 
\begin{align}\label{eq:ric}
\begin{split}
i\hbar\vv_f\grad \hat\gamma + 2\epsilon \hat\gamma = \hat\gamma \underline{\hDelta} \hat\gamma 
- \hat\Delta \,,
\\
i\hbar\vv_f\grad \hat{\underline\gamma} - 2\epsilon \hat{\underline\gamma} 
= \hat{\underline\gamma} \hat\Delta \hat{\underline\gamma} 
- \underline{\hDelta} \,.
\end{split}
\end{align}
These functions carry information about particle-hole coherence 
and can be expressed through Andreev amplitudes, for example $\hat\gamma = \hat u^{-1} \hat v$
\cite{Eschrig:2000ux}. 
For retarded functions ($Im(\epsilon) > 0$), 
the integration of these equations are done along the straight classical trajectory 
in direction of Fermi velocity $\vv_f$ 
for $\hat\gamma$, and in $-\vv_f$ direction for $\hat{\underline\gamma}$. 
In unitary superfluids 
$\hat\Delta \underline\hDelta = - |\Delta(\vp_f)|^2 \hat{1}$, 
solution for amplitudes in uniform state are:
\begin{equation}
\hat\gamma(\vp_f;\epsilon) 
= -\frac{\hat\Delta(\vp_f)}{\epsilon \pm i\sqrt{|\Delta(\vp_f)|^2 - \epsilon^2}}
\,,\qquad
\hat{\underline\gamma}(\vp_f;\epsilon) 
= \frac{\underline\hDelta(\vp_f) }{\epsilon \pm i\sqrt{|\Delta(\vp_f)|^2 - \epsilon^2}}
\,,
\label{eq:Riccati_uniform}
\end{equation}
with the signs distinguishing between positive/negative $Im(\epsilon)$. These solutions are 
taken as initial values for the coherence amplitudes far away from the interface. 

Integration of the transport equations 
require 
boundary conditions for the propagators, or for the coherence amplitudes. 
There exist several models that are based on different physical pictures 
of the scattering process at a surface. For atomically smooth 
surface the parallel momentum is conserved and the reflection is mirror-like, or specular. 
In this case both the 
propagator and the coherence amplitudes are continuous across the reflection point. 
For atomically rough surfaces one can use 
`randomly rippled wall' model \cite{BUCHHOLTZ:1979wr,BUCHHOLTZ:1986tc,BUCHHOLTZ:1991ts}, 
diffuse boundary with thin layer of atomic-size impurities coating a smooth surface
\cite{Kopnin:1986cv,ZHANG:1987vq},
`randomly oriented mirror' model \cite{THUNEBERG:1992us}, 
and a universal model based on scattering S-matrix approach that 
can describe partial specular-diffuse reflection\cite{Nagato:1996vb,Nagato:1998we}. 
Typically, various implementations of boundary conditions give similar results for 
the order parameter but somewhat different quasiparticle spectra. 
Most recent formulation of the boundary conditions is due to M. Eschrig, 
who described an S-matrix approach to scattering in terms of coherence amplitudes\cite{Eschrig:2009ub}. 

\begin{figure}[t]
\centering\includegraphics[width=2.5in]{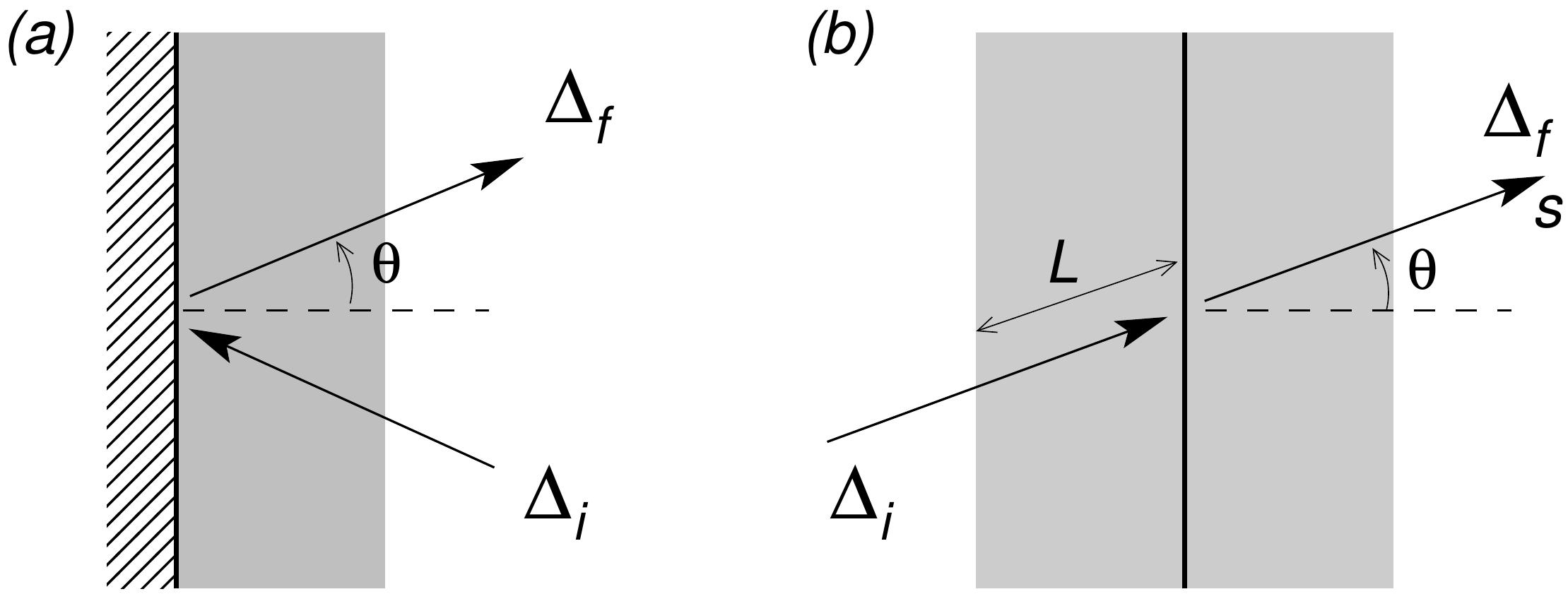}
\caption{The quasiclassical trajectories at a specular surface or 
completely transparent domain wall. 
The propagator at the surface/interface is determined by the coherence amplitudes 
$\hat\gamma_i$ integrated along the incoming trajectory, and 
$\hat{\underline\gamma}_f$ integrated opposite the outgoing trajectory (retarded functions). 
The order parameter configurations along two trajectories 
far away from the interface determine the `topological' aspect of 
the bound states; whereas the region of suppressed order parameter 
of typical lengths 
$L \sim 5\xi_0/\cos\theta$ along a trajectory, 
determines multiple-reflected quasiparticle states in effective order parameter potential well. 
The coherence length is defined as $\xi_0 = \hbar v_f/ 2\pi k_B T_c$. 
}
\label{fig:absdw}
\end{figure}

Using Riccati amplitude 
parametrization of the Green's functions it is particularly straightforward 
to estimate the local density of states (DOS) and obtain structure of 
the bound states spectrum at specular boundary 
or completely transparent interface, \figref{fig:absdw}. 
The spin-matrix density of states is defined as 
$\hat N(\vR, \vp_f; \epsilon) = -(1/\pi) Im \, \hg(\vR, \vp_f; \epsilon)$
and at the surface is determined by:
%
\begin{equation}
\hat{g}^{\scriptscriptstyle R}(\vR=0, \vp_f; \epsilon) 
= -i \pi \frac{1 + \hat\gamma(0,\vp_f;\epsilon) \hat{\underline\gamma}(0,\vp_f;\epsilon)}
{1 - \hat{\gamma}(0,\vp_f;\epsilon) \hat{\underline\gamma}(0,\vp_f;\epsilon)}
\end{equation}

Neglecting the suppression of the order parameter near the interface, 
we can approximate the coherence amplitudes by their uniform values 
Eq.~(\ref{eq:Riccati_uniform})
far from the interface in final and initial points of a trajectory: 
$\hat{\gamma}(0,\vp_f;\epsilon)=\hat{\gamma}_i$ and
$\hat{\underline\gamma}(0,\vp_f;\epsilon)=\hat{\gamma}_f$, 
where we assume the coherence amplitude is continuous along the 
trajectory at $\vR=0$, thus the same value of the momentum $\vp_f$. 

In singlet superconductors with one OP component this leads to condition 
on the poles of the Green's function $\hat{g}(0)$ at the surface 
\be
1 -\hat{\gamma}_i \hat{\underline\gamma}_f 
= 1 - \frac{ \Delta_i \Delta_f^*}{
[\epsilon+i\sqrt{|\Delta_i|^2 - \epsilon^2}]
[\epsilon+i\sqrt{|\Delta_f|^2 - \epsilon^2}]
} = 0 \,.
\ee
The spectrum of bound states for parametrization 
$\Delta_i = \Delta e^{+i\varphi/2}$, 
$\Delta_f = \Delta e^{-i\varphi/2}$ is:
\be
1 - \left[ \frac{ \Delta e^{+i\varphi/2} }{
\epsilon+i\sqrt{\Delta^2 - \epsilon^2} }
\right]^2 = 0
\qquad\Rightarrow\qquad
\epsilon = \pm \Delta \cos\frac{\varphi}{2}
\ee
which gives well-known zero-energy states for trajectories 
that connect order parameter values with relative phase $\pi$ \cite{Hu:1994vo}. 

As another example, we consider interface between two degenerate states 
of \He-B. The two states have the same 
energy gap, but generally have order parameter that has an invariant 
part $\vDelta_+$ and a sign-changing part $\vDelta_-$ on two sides: 
$\vDelta_i = \vDelta_+ - \vDelta_-$, 
$\vDelta_f = \vDelta_+ + \vDelta_-$, with 
$\vDelta_i^2 = \vDelta_f^2 = \vDelta_+^2 + \vDelta_-^2 \equiv \Delta^2$.
The denominator of the diagonal propagator $\hg(0)$ is 
\be
1 -\hat{\gamma}_i \hat{\underline\gamma}_f 
= 1 - \frac{ (\vDelta_i\vsigma) (\vDelta_f \vsigma)}{
[\epsilon+i\sqrt{\Delta^2 - \epsilon^2}]^2
} 
= 1 - \frac{ \vDelta_i \cdot \vDelta_f + i \vsigma (\vDelta_i \times \vDelta_f)}{
[\epsilon+i\sqrt{\Delta^2 - \epsilon^2}]^2
} 
= 0 \,.
\ee
The spin structure of the bound states is determined by cross product 
of spin vectors in final and initial points of the trajectory, 
$\propto (\vDelta_i \times \vDelta_f)$, 
and the energies of the bound states follow from poles of 
$(1 -\hat{\gamma}_i \hat{\underline\gamma}_f )^{-1}$: 
\be
\left[ 
1 - \frac{ \vDelta_+^2 - \vDelta_-^2 }{ (\epsilon+i\sqrt{\Delta^2 - \epsilon^2})^2}
\right]^2 + 
\left[
\frac{  2 \vDelta_+ \times \vDelta_-}{ (\epsilon+i\sqrt{\Delta^2 - \epsilon^2})^2 }
\right]^2 =0 \,.
\ee
After re-arranging this gives bound state energies:
\be
2 \epsilon^2 - 2 \vDelta_+^2 + 2i\epsilon\sqrt{\Delta^2 - \epsilon^2} = \pm i 2 |\vDelta_+| |\vDelta_-|
\qquad\Rightarrow\qquad
\epsilon = \pm |\vDelta_+| \,,
\ee
- determined by the order parameter component that remains 
invariant in the reflection/transmission process.
\begin{figure}[t]
\centering\includegraphics[width=0.48\linewidth]{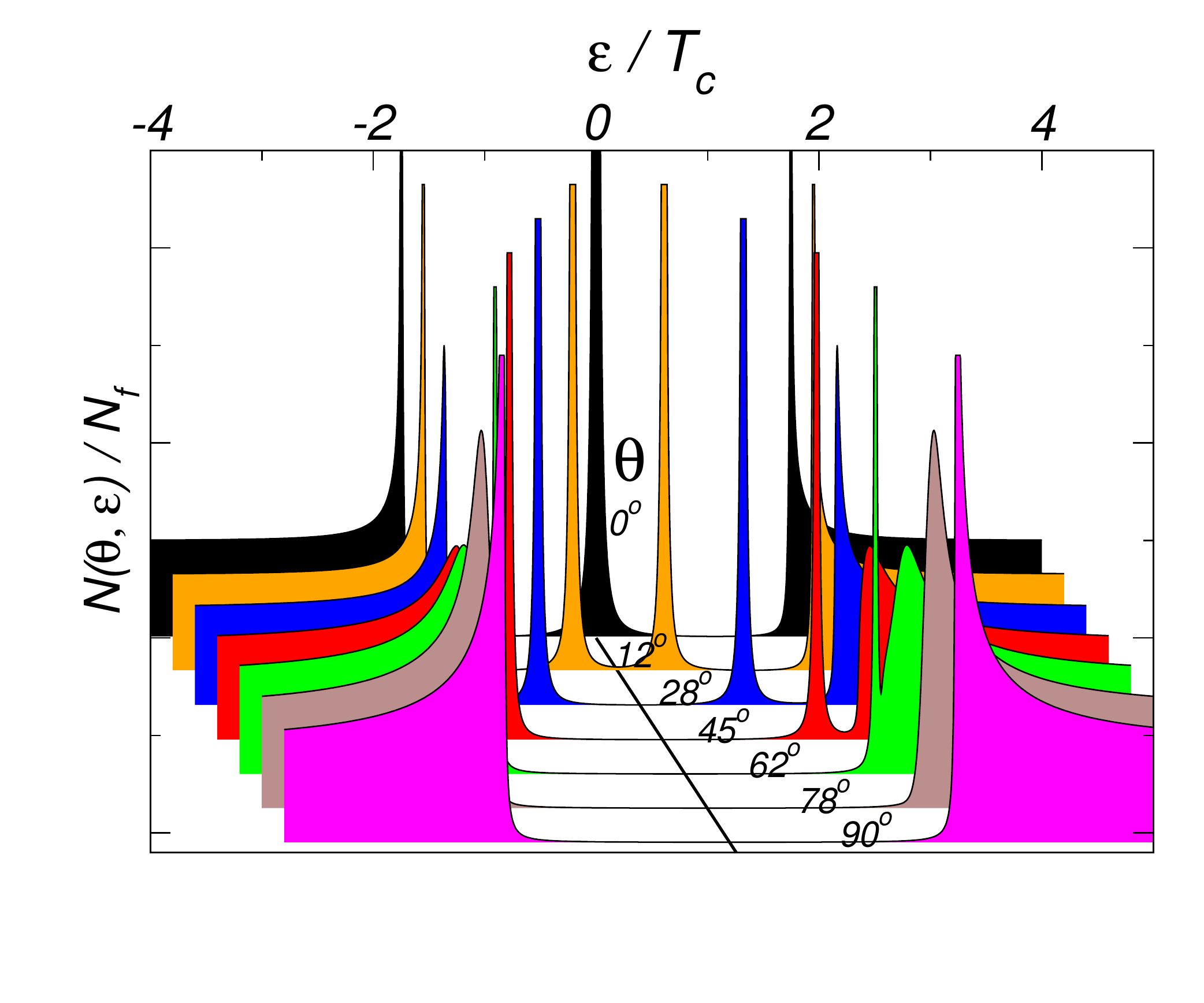}
\centering\includegraphics[width=0.48\linewidth]{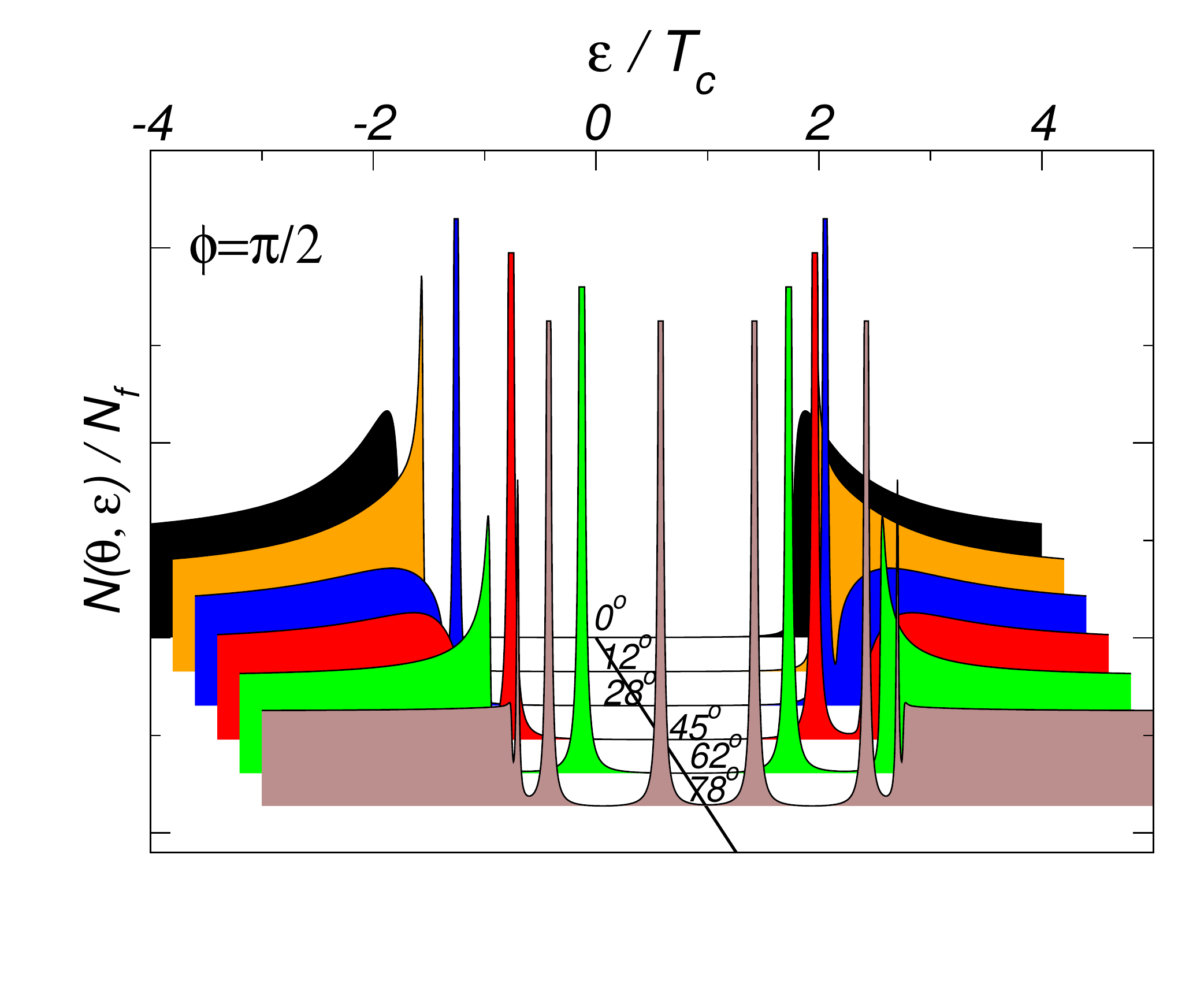}
\caption{The angle-resolved DOS in domain walls between degenerate 
states of superfluid \He-B. 
On the left, quasiparticles that travel across domain wall (transverse direction), 
experience OP sign change. 
This configuration also gives spectrum at a specular surface. The bound states 
form a cone in momentum-energy space Eq.~(\ref{eq:dwBbsZ}).
In the right panel, 
one of the parallel OP components ($y$ in this case) changes sign, 
resulting in bound states near continuum edge, shown here for $\hp_x=0$, Eq.~(\ref{eq:dwBbsY}). 
}
\label{fig:dwBdos}
\end{figure}
We use it evaluate DOS for two domain wall configurations that are relevant to the 
film geometry. Plane $xy$ separates two domains at $z<0$ and $z>0$. 
In the first configuration the `normal' OP component changes sign:
from $\vDelta_i(z=-\infty) = (\Delta_0 \hp_x, \Delta_0 \hp_y, -\Delta_0 \hp_z)$ to
$\vDelta_f(z=+\infty) = (\Delta_0 \hp_x, \Delta_0 \hp_y, +\Delta_0 \hp_z)$, 
resulting in the bound states energies 
\be
\epsilon_{\perp} = \pm \Delta_0 \sqrt{\hp_x^2 + \hp_y^2} = \pm \Delta_0 \sin\theta \,.
\label{eq:dwBbsZ}
\ee
In second configuration, quasiparticle that travel `parallel' to the wall experience 
the sign change, 
from $\vDelta_i(z=-\infty) = (\Delta_0 \hp_x, -\Delta_0 \hp_y, \Delta_0 \hp_z)$ to
$\vDelta_f(z=+\infty) = (\Delta_0 \hp_x, +\Delta_0 \hp_y, \Delta_0 \hp_z)$, 
and form bound states with 
\be
\epsilon_{\parallel} = \pm \Delta_0 \sqrt{\hp_z^2 + \hp_x^2} 
= \pm \Delta_0 \sqrt{ \cos^2\theta + \sin^2\theta \cos^2\phi} \,.
\label{eq:dwBbsY}
\ee
The angle resolved DOS for the two domain walls from self-consistent numerical calculation is 
shown in \figref{fig:dwBdos}. The approximate calculation of bound states energies 
agrees with it quite well. 
The bound states in `parallel' configuration in general lie closer to the 
continuum states and result in a lower energy for this domain wall 
(see also section \ref{sec:new}\ref{sec:newHe}). 

The above states arize from the `topological' properties of the 
particular domain wall or surface orientation. 
To include other bound states that appear as 
a result of multiple particle-hole reflections inside the OP suppression region, 
or to find DOS at arbitrary distance from the surface, one should know the 
coherence amplitudes everywhere along a trajectory. For a piecewise order parameter profile 
this can be done using a property of Riccati equations. 
In a region with $\hat\Delta_0=const$ and known uniform solution $\hat\gamma_0$, 
solution for $\hat\gamma$ with (another) initial value $\hat\gamma(0)$ has the 
form
$\hat\gamma = \hat\gamma_0 + \hat{z}^{-1}$
with auxiliary function 
$\hat z$ having initial value $\hat{z}(0) = (\hat\gamma(0) - \hat\gamma_0)^{-1}$ 
and satisfying linear equation that is more readily solved than non-linear one:
\be
-i\hbar\vv_f\grad \hat{z} + 2\epsilon \hat{z} = 
\underline{\hDelta}_0 \hat\gamma_0 \hat{z} +
\hat{z} \hat\gamma_0 \underline{\hDelta}_0 
+ \underline{\hDelta}_0 \,.
\ee

Finally, to determine the relative stability of different phases, one needs to calculate 
the free energy, given an OP configuration. This can be done 
using the Eilenberger functional\cite{Eilenberger:1968wb}, or one of  
the approaches based on the Luttinger-Ward functional with differentiation with respect to 
coupling constant, as applied to thin films of \He-A\cite{VorontsovAB:2003ki}, 
or differentiation with respect to the energy, 
as described in application to multi-order pnictide materials \cite{Vorontsov:2010tk}.

\section{New superfluid phases in confined geometry}
\label{sec:new}

Within about $5-10\xi_0$ of a pairbreaking surface, the order parameter is strongly 
suppressed, and a re-distribution of the 
quasiparticle spectrum takes place. States from continuum are shifted below the 
gap and appear as bound states. 
As the spectrum is modified, significant changes occur to the thermodynamic and transport 
properties in this region. If a superconducting state is confined to a slab 
only two to three times larger than $5-10\xi_0$ then the pairbreaking influence of the surfaces 
extends over the entire volume of the sample and one expects that the 
changes in properties of superconducting state will be detectable. 
Some of the changes constitute a somewhat `trivial' modification of the superfluid phases, 
that does not strongly affect the overall structure and symmetry of the 
condensate. 
Despite this, such modifications are interesting in themselves from another perspective: 
Andreev bound states reflect the non-trivial  
topological aspects of the underlying superfluid phases, 
as shown, for example, by investigating 
angular momentum in discs of chiral superfluid\cite{Sauls:2011tc,Tsutsumi:2012uw},
or superflow in narrow channels\cite{Tsutsumi:2011eo,Wu:2013ev}. 

The other, more dramatic effect of strong confinement and 
large overall suppression of order parameter, 
is the possibility of appearance of new phases with different symmetry 
properties compared with the bulk phases. Andreev bound states also 
play the key role here, as the new order parameter configurations and new 
broken symmetries 
are determined by the changes in energies of these states. 

Prediction of appearance, and investigation of new phases in confined geometry usually 
falls outside the scope of the traditionally employed GL theory, 
that has only lowest gradient terms, $|\grad \Delta|^2 \ll |\Delta^2|/\xi_0^2$. 
The strong confinement is associated with fast changing order 
parameter and large influence of the gradient energy terms in the free energy functional. 
This indicates a great importance of non-local effects for these phenomena. 
In nodal superconductors non-locality can play especially important role 
since coherence length along nodal directions is very long, 
resulting, for example, in a modified temperature behavior of 
penetration depth.\cite{Kosztin:1997tl}

These non-local effects can be interpreted as interaction of Andreev bound states 
across the width of the slab, and can result in self-generation of 
bound states and currents in a semi-infinite material. 
The new superfluid states that appear in constrained environment have  
lower symmetry than allowed by geometry. Nature of the new broken symmetry 
depends on the properties of the pairing interaction, its symmetry and number of 
order parameter components, 
as well as the background superfluid phases. 
Below we describe two examples, where the additional broken symmetries 
are time-reversal symmetry and spatial translation symmetry.

\subsection{Singlet superconductors}
\label{sec:newD}

\begin{figure}[t]
	\centerline{\includegraphics[width=0.3\linewidth]{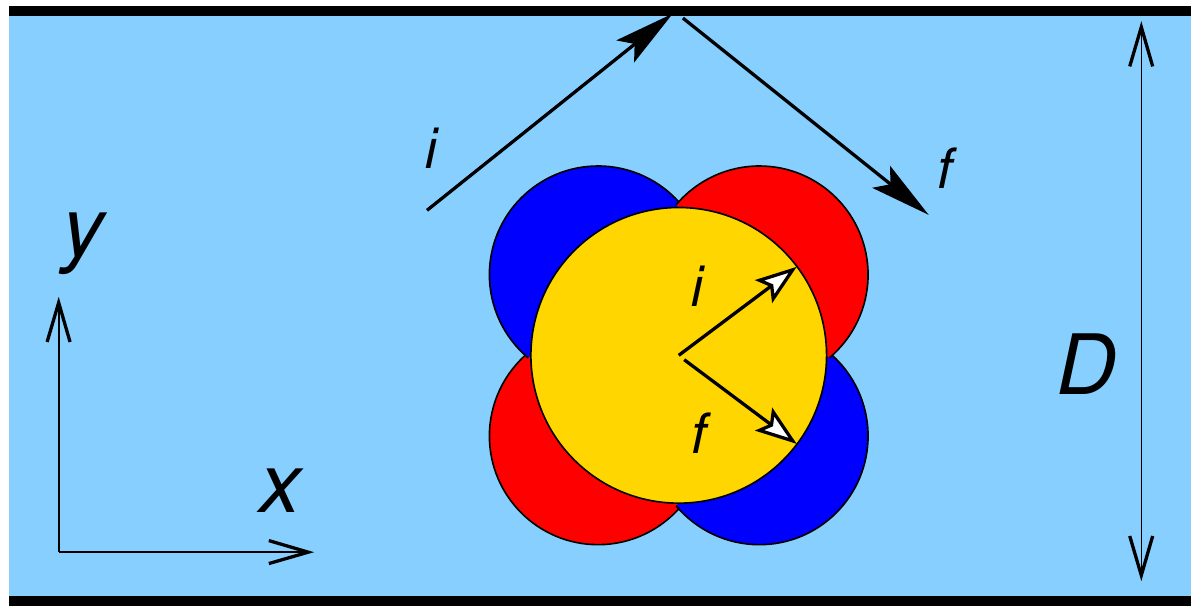}}
\caption{
	A superconducting wire/slab of thickness $D$ with strongly pairbreaking surfaces. 
}
\label{fig:dwire}
\end{figure}

We first discuss confinement of a one-component singlet $d$-wave superconductor between 
two specular pairbreaking surfaces, \figref{fig:dwire}, where the order parameter 
is $\Delta(\vR, \vp_f) = \Delta(y) \cY(\vp_f)$ and $\cY(\vp_f) = \sin2\phi_p$ 
is the basis function ($\tan\phi_p = \hp_y/\hp_x$). 
This simple system clearly demonstrates effects of confinement on Andreev bound states and the 
role they play in generating phases with additional broken symmetries. 

\begin{figure}[t]
	\includegraphics[width=0.48\linewidth]{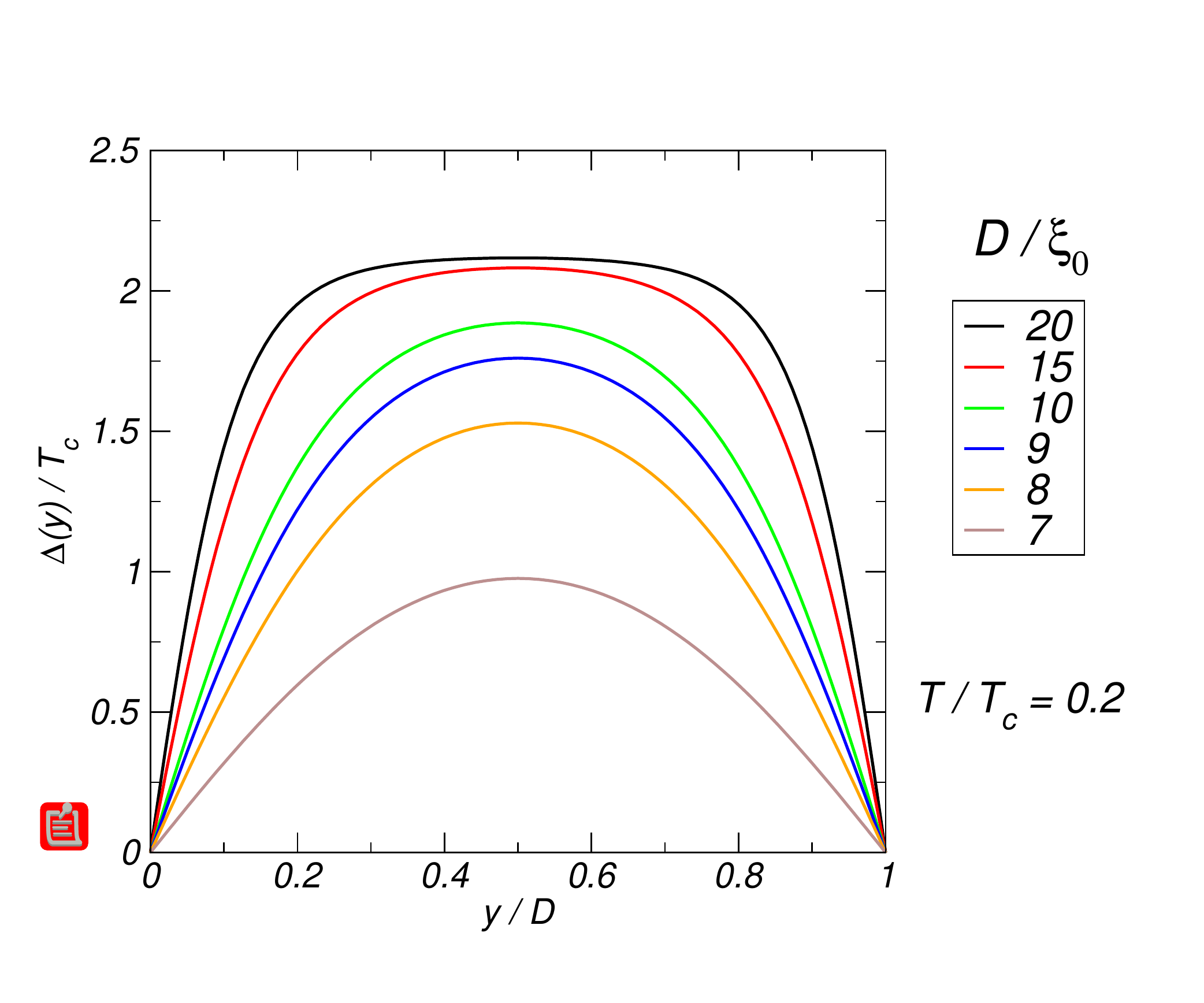}
	\hfill
	\includegraphics[width=0.48\linewidth]{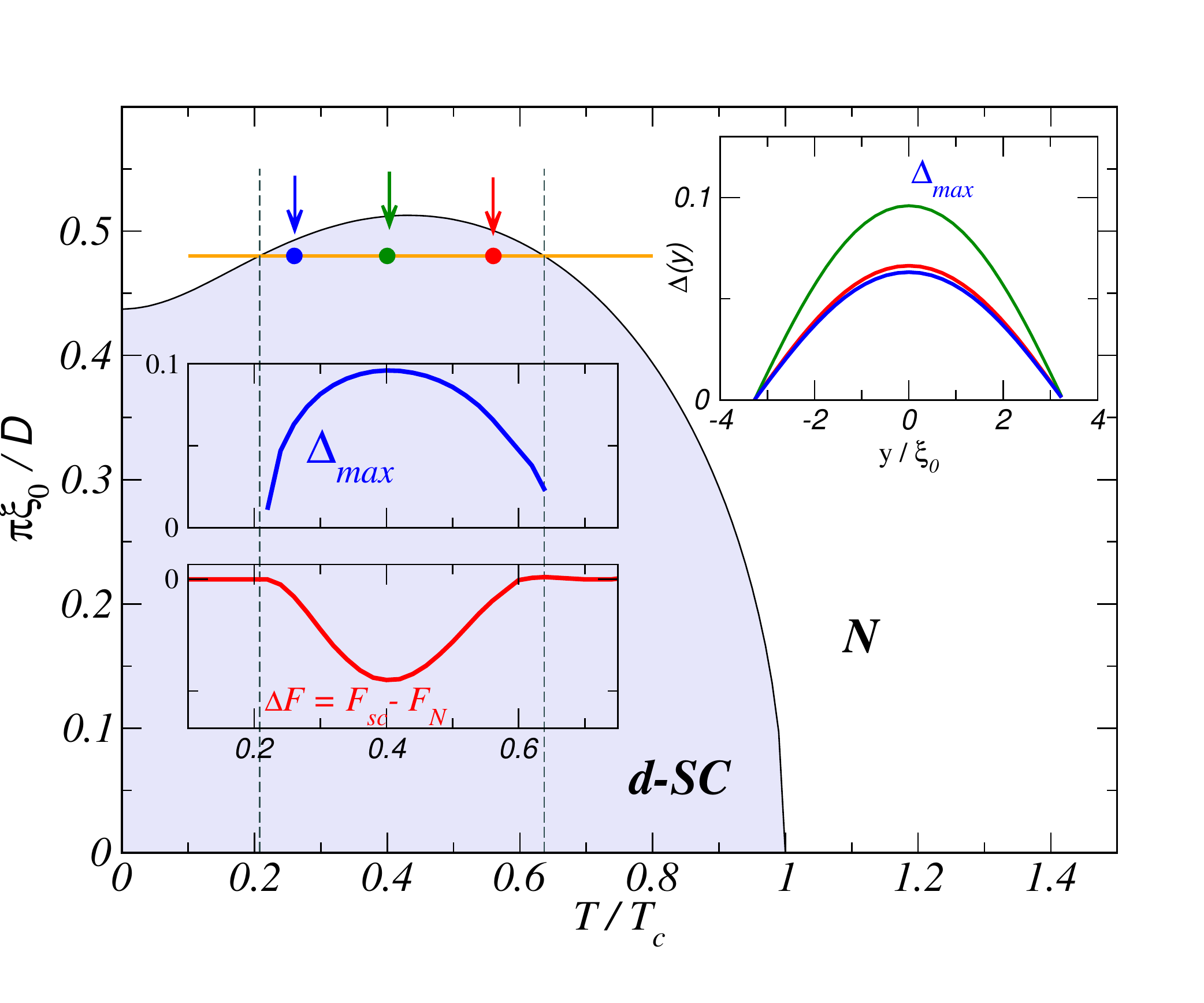}
\caption{
	The normal-superconducting (N-SC) transition in $d$-wave films. 
	Left: gradual suppression of the order parameter in the 
	film with decreasing thickness $D$ at fixed temperature. 
	Right: the ``temperature-confinement'' phase diagram of 
	a $d$-wave film. The N-SC transition line shows 
	re-entrant behavior. The insets present the amplitude 
	of the order parameter and free energy along the line traversing the 
	re-entrance region, and profiles of the 
	order parameter at three points along the line, indicated by the color-coded arrows. 
}
\label{fig:dwavefilm}
\end{figure}

We start by looking at the confinement-induced transition between superconducting 
and normal phases. A complete description of the superconducting state requires 
self-consistent calculation of the non-uniform order parameter. 
Spatial profiles of the gap function $\Delta(y)$ for films of various 
slab thicknesses $D$, obtained using 
quasiclassical theory, are shown on the left in \figref{fig:dwavefilm}. 
As the slab becomes narrower, superconductivity gets suppressed and disappears 
below about $D \sim 7\xi_0$ (for temperature $T/T_c =0.2$). 

It is convenient to introduce a parameter measuring confinement, or 
inverse thickness, $\tilde Q \equiv \pi \xi_0/D$, and plot the N-SC transition 
in confinement vs temperature plane, shown on the right of \figref{fig:dwavefilm}. 
When $\tilde Q = 0$ the system is infinite, or semi-infinite, transition $T_c$ is that of the bulk 
system. As confinement increases the transition temperature is suppressed. 
However, in thin films there is a range of re-entrance 
of the normal phase, $0.44 \lesssim \tilde Q \lesssim 0.51$, see also \cite{Nagato:1995tu}.
If one traverses this region toward lower $T$, the order parameter first appears and grows, 
but then starts to drop, smoothly disappearing at re-entrance of the normal state. 
The free energy is lower than the normal state value along this path, 
as shown in the insets of the right panel of \figref{fig:dwavefilm}.

To understand details of this transition, why back-bending feature 
appears and whether it is physical, 
one should derive a free energy functional expansion in small order parameter $\Delta$, 
keeping gradient energy terms to all orders, since they play the main role, as we will see. 
Expanding the order parameter into plane wave basis
$$
\Delta(\vR, \vp_f) =  \cY(\vp_f) \sum_\vq \, \Delta_\vq e^{i\vq\vR} \,
$$
we can integrate back the self-consistency equation to obtain 
the free energy functional:
\be
\frac{\delta \Del F}{\delta \Delta^*_\vq} \equiv 
\Delta_\vq \, \left< \cY(\vp_f)^2 \right> \ln {T\over T_c} 
- T\sum_{\epsilon_m}  \left< \cY(\vp_f) 
\left( f_\vq (\vp_f; \epsilon_m) - {\pi \Delta_\vq \cY(\vp_f) \over |\epsilon_m|} \right) 
\right> =0 \,.
\label{eq:selfc}
\ee 
Here 
$\epsilon_m= \pi T (2m+1) $ - Matsubara energy, 
and brackets denote angle average over circle Fermi surface, 
$\langle \dots \rangle = \int \frac{d{\phi_{\vp}}}{2\pi} \dots $.
To get the free energy, we need to solve the quasiclassical equations 
to third order in $\Delta_\vq$. The diagonal, $g$, and off-diagonal (anomalous) components, 
$f$ and $\underline{f}$, 
of the quasiclassical Green's function 
are found from the set of equations: 
\begin{align}
\begin{split}
& [\epsilon_m + \frac12\vv_f \cdot\grad] f (\vR, \vp_f; \epsilon_m) 
= i\,g \; \Delta(\vR, \vp_f)  \,,
\\
& g^2 -f \underline{f} = -\pi^2 \,, \quad
\underline{f}(\vR, \vp_f; \epsilon_m)  = f(\vR, \vp_f; -\epsilon_m)^* \,.
\end{split}
\label{eq:eil}
\end{align}
Following the step-by-step perturbation scheme 
one obtains the free energy functional \cite{VorontsovAB:2009ef}:
\begin{align}
\label{eq:fe}
\begin{split}
& \Del F = \sum_\vq \; I(T, \vq) \, |\Delta_\vq|^2 + 
{1\over 2} \sum_{\vq_1+\vq_2 = \vq_3+\vq_4} \, K(T, \vq_1, \vq_2, \vq_3, \vq_4) \;
\Delta_{\vq_1}^* \Delta_{\vq_2}^* \Delta_{\vq_3} \Delta_{\vq_4} \,,
\\
& I(T,\vq) = \left< \cY(\vp_f)^2 \right> \ln {T\over T_c} - 2\pi T \sum_{\epsilon_m>0} Re 
\Big< \cY^2(\vp_f) \left(  {1\over \epsilon_m + i \eta_{\vq}} - {1\over \epsilon_m} \right) \Big> \;,
\\
& K(T, \vq_1, \vq_2, \vq_3, \vq_4) = 2\pi T \sum_{\epsilon_m>0} {1\over2} Re 
\Big< 
\cY^4(\vp_f) \frac{\epsilon_m+i(\eta_{\vq_1}+\eta_{\vq_2}+\eta_{\vq_3}+\eta_{\vq_4})/4}
{(\epsilon_m+i\eta_{\vq_1}) (\epsilon_m+i\eta_{\vq_2}) (\epsilon_m+i\eta_{\vq_3}) (\epsilon_m+i\eta_{\vq_4})}
\Big> \,.
\end{split}
\end{align}
where $\eta_\vq = \frac12 \vv_f \cdot \vq $. 

The transition from normal to superconducting state is determined by 
vanishing quadratic coefficient, $I(T,\vq)=0$. If we take a uniform state along the 
slab the order parameter is vanishing at the specular surfaces $y=\pm D/2$, and has the 
form $\Delta(y) =\Delta \cos Qy$ with $Q=\pi/D$, and the modulating vectors are $\vq = (0, \pm Q)$. 
After angle integration over cylindrical Fermi surface, 
\be
2 I(T, \eta=\frac12 v_f Q)  = \ln\frac{T}{T_c} - 2\pi T \sum_{\epsilon_m>0} 
\left[ \frac{4\epsilon_m^4}{\eta^4} \left( \sqrt{1+\frac{\eta^2}{\epsilon_m^2}}-1 \right)^2 - 1    
\right] \frac{1}{\epsilon_m}
\ee
The exact numerical solution for the instability is presented in
\figref{fig:dwavefilm}. At large $Q$ it has the re-entrant feature, where 
upon cooling the superconducting order appears and then disappears again 
into the normal phase. 
To qualitatively understand this behavior we can take the sum in 
the limit $\eta \ll \pi T$: 
\be
\Del F = \frac12 \Delta^2 \left[ 
\ln\frac{T}{T_c} 
+ \frac{7\zeta(3)}{8} {\tilde Q}^2 \frac{T_c^2}{T^2} 
\right]
+ \mathcal{O}(\Delta^4)
\ee
The temperature-dependent non-local term, arising from the gradient 
energy, is the one responsible for the re-entrance. 
At certain fixed $Q$, the instability equation 
$I(T,\eta) = 
\ln\frac{T}{T_c} 
+ \frac{7\zeta(3)}{8} {\tilde Q}^2 \frac{T_c^2}{T^2} =0$
can have two solutions $T_{high}$ and $T_{low}$, that 
means that at the $T_{high}$ the negative 
log term dominates and order parameter appears, 
but as temperature is lowered, the non-local 
positive term takes over and superconductivity is 
suppressed $T_{low}$. 
The order parameter is changing smoothly in this range, 
and the free energy is lower than that of the normal phase. 
The approximation $\xi_0 Q \ll T/T_c$ gives only the qualitative 
picture; a more careful treatment of the first, $m=0$, 
non-linear term leads to the same conclusion and allows 
for more precise determination of the N-SC transition line 
in the limit $\tilde Q \approx \pi T/T_c$. 

Another explanation of a similar re-entrance feature 
in a single-component polar state $\Delta(y) \hat{p}_y$ 
confined to a slab, 
was presented in \cite{Hara:1986dn}.
It was shown that it is a 
direct consequence of the midgap states and dependence of their energy spectrum 
on the thickness of the slab. 

One of the important features, relevant to the re-entrant N-SC transition, 
is that the fourth-order coefficient 
$K$ is always positive in 
the region close to the second-order instability. 
Because of this, the smooth variation of the order parameter with temperature 
results in a smooth variation of the free energy. 
The entropy in this temperature range can be less or 
greater than the normal state's value 
$\Del S (T) = -\partial \Del F /\partial T$, but 
the overall entropy $S_n(T) + \Del S(T)$ has positive temperature 
derivative (positive specific heat), and  
no thermodynamic inequality is violated in the re-entrant regime. 
This can be contrasted with back-bending feature of linearized gap equation 
of paramagnetic depairing, where $K(T)$ changes sign along N-SC transition line, 
resulting in a first-order transition at lower temperatures. 
It was argued that the back-bending feature in that case indicates 
thermodynamically unstable configuration due to entropy decrease 
\cite{jam68}. 

Further analysis of the free energy in linearized regime reveals that 
the translation-invariant state in the film is 
not stable at low temperatures \cite{VorontsovAB:2009ef}. 
The state that is realized in constrained geometry is one that 
carries current and breaks time-reversal symmetry. The OP structure and the free 
energy of current-carrying state close to the N-SC transition is: 
\be
\Delta(\vR) = \Delta_2 e^{i Q_x x} \cos Q_y y  \,,
\quad \mbox{} \quad
\Del F[Q_x] = -\frac{2 \; I^2(T,\vQ)}{2(K_1+2 K_{12}) } \,,
\label{eq:FEb}
\ee
\label{eq:FE}
denoting $K_1 = K(\vq_1,\vq_1,\vq_1,\vq_1)$, 
$K_{12} = K(\vq_1,\vq_2,\vq_1,\vq_2)$. 
The other possible state that involves linear combination of 
order parameters with two opposite wave vectors, or amplitude 
modulation of the OP, loses in energy to the current-carrying state \cite{VorontsovAB:2009ef}.

\begin{figure}[t]
\begin{minipage}{0.45\linewidth}
\centering\includegraphics[width=1.2\linewidth]{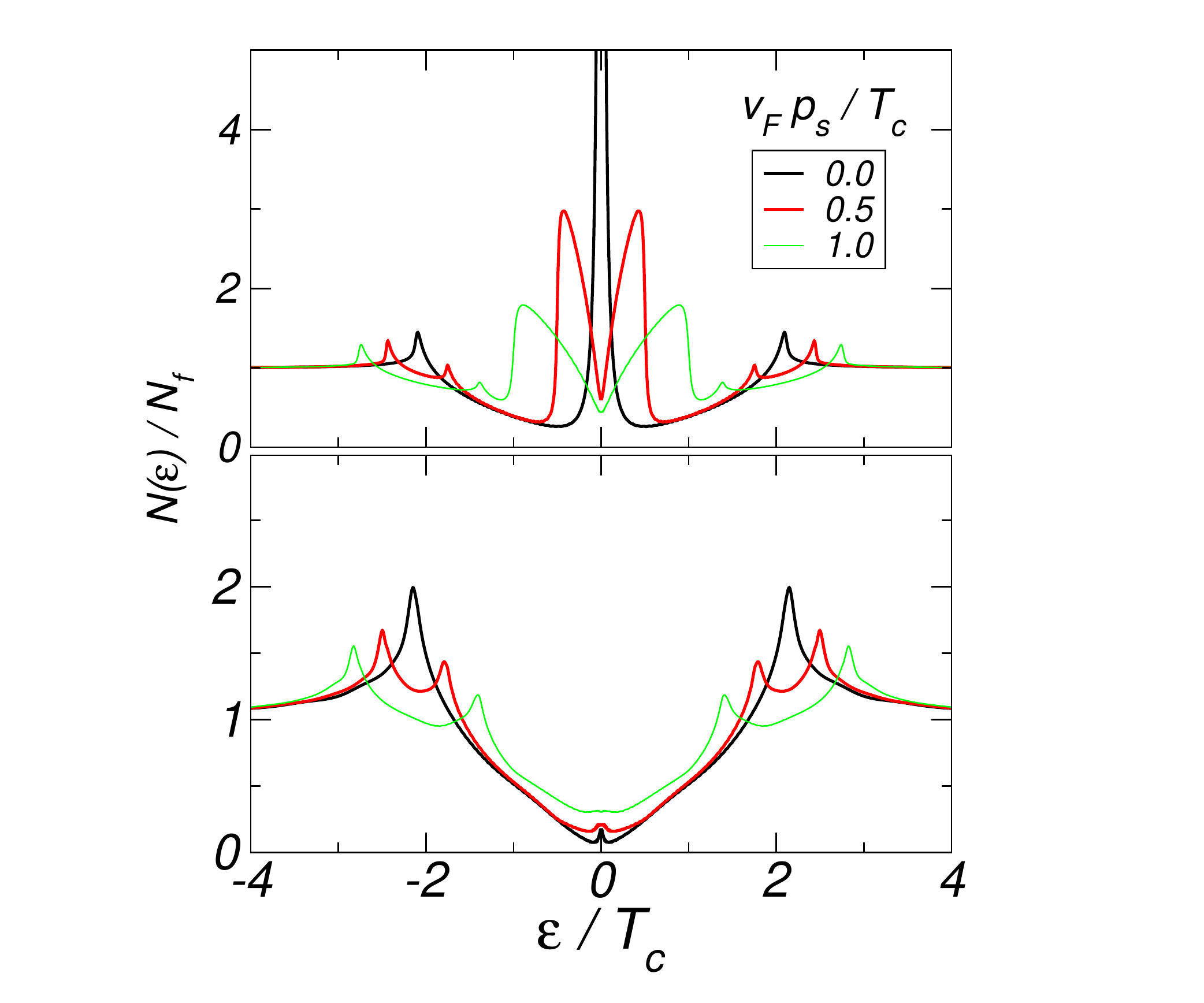}
\end{minipage}
\begin{minipage}{0.55\linewidth}
\centering\includegraphics[width=1.1\linewidth]{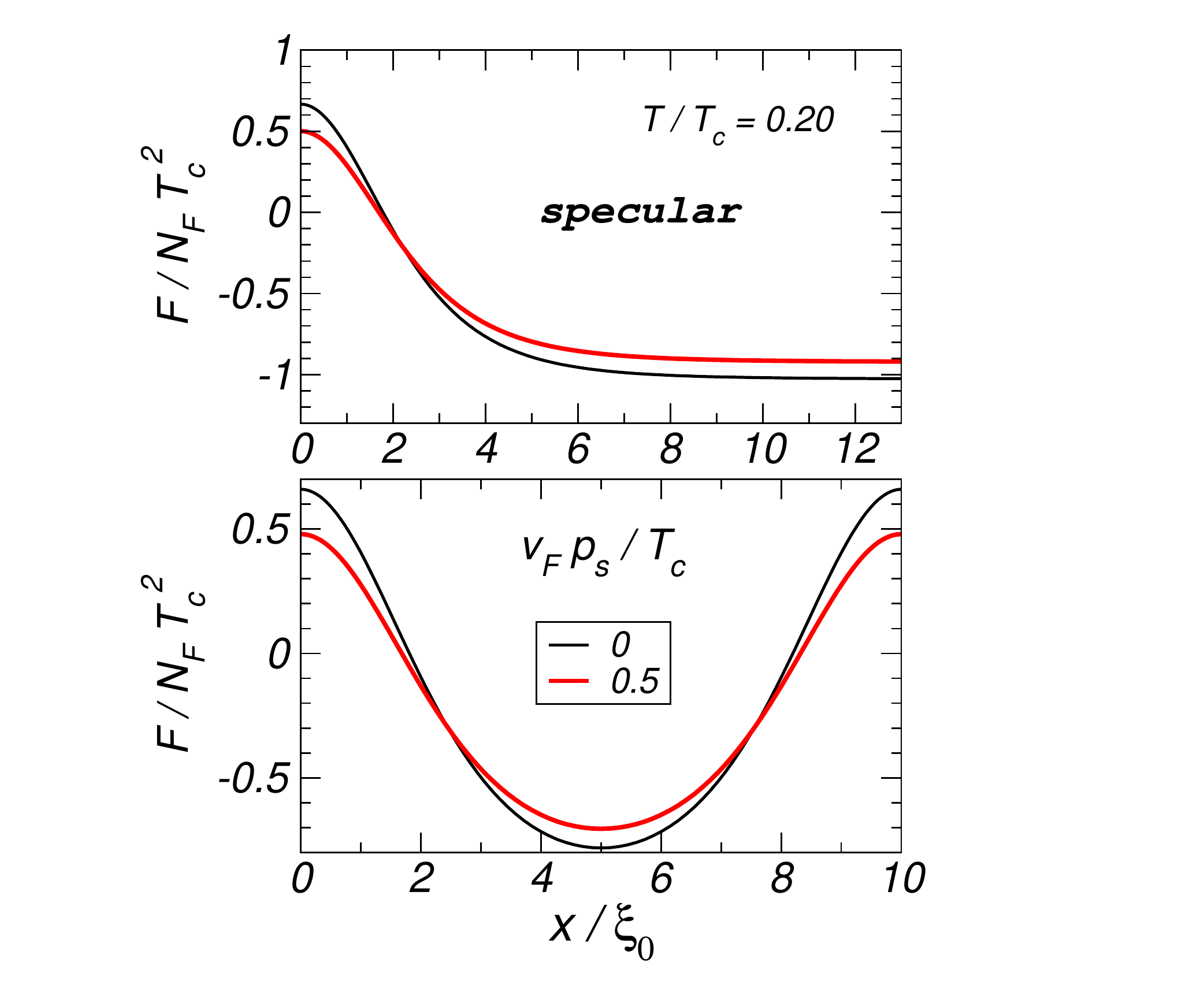}
\end{minipage}
\caption{
	Left: density of states in a semi-infinite $d$-wave superconductor. 
	The zero-energy peak of Andreev bound states at a specular surface 
	is split by superflow (top left), whereas far away from the surface 
	there is no zero-energy states (bottom left). 
	Right: free energy profile. Bound state splitting lowers free energy at the surface, 
	and continuum states dominate bulk region $x \gtrsim 4 \xi_0$, 
	increasing free energy (top right). 
	In thin films the surface gain in energy due to split ABS can overcome energy loss 
	in the center (bottom right), favoring current-carrying state. 
}
\label{fig:dwall}
\end{figure}

The reason for stability of the current-carrying state 
that breaks time-reversal symmetry 
is Andreev bound states near pairbreaking surfaces of the film. 
These states are known to carry paramagnetic current and they result in the 
lowering of surface energy.
One can see this by looking at the free energy difference between 
a state with superflow $\vp_s$ and time-reversal invariant state, $p_s = 0$, expressed in 
terms of the local density of states: 
\be
F(\vR,\vp_s) - F(\vR,0) = - 2N_f \int\limits_{-\infty}^{+\infty} \frac{d\epsilon}{2}
\left<
2T \ln\left[2\cosh\frac{\epsilon}{2T}\right] \left( N_{\vp_s}(\vR,\vp_f; \epsilon) -
N(\vR, \vp_f; \epsilon)\right)
\right> \,.
\label{eq:FEps}
\ee
This expression is valid when the superflow is small, i.e. when pairbreaking 
effects of the superflow on the order parameter can be neglected. 
This way the presence of 
a current is only reflected via the Doppler shift of the energy spectrum: 
$N_{\vp_s}(\vR, \vp_f; \epsilon) = N(\vR, \vp_f; \epsilon - \vp_s(\vR)\vv_f) $.
\footnote{
This form of the free energy can be obtained from the Luttinger-Ward functional, 
and one can verify that it gives correct expression for local current in terms of the 
density of states,
$$
\vj_s(\vR) \equiv \pder{F(\vR,\vp_s)}{\vp_s} = 
2 N_f \int\limits_{-\infty}^{+\infty} \frac{d\epsilon}{2}
\left< \vv_f 
2T \ln\left[2\cosh\frac{\epsilon}{2T}\right] 
\pder{}{\epsilon} N(\vR, \vp_f; \epsilon-\vp_s(\vR)\vv_f)
\right>
$$
$$
=
- 2N_f \int\limits_{-\infty}^{+\infty} \frac{d\epsilon}{2}
\tanh\frac{\epsilon}{2T}
\left< \vv_f 
N( \vR, \vp_f; \epsilon-\vp_s(\vR)\vv_f)
\right>
$$
where in the last step we did integration by parts.
This agrees with the usual definition of the current 
$
\vj_s = 
2N_f \int\limits_{-\infty}^{+\infty} \frac{d\epsilon}{4\pi i}
\left< \vv_f  \; g^K
\right>
$
after we write the Keldysh propagator in equilibrium as 
$
g^K=
(g^R - g^A) \tanh\frac{\epsilon}{2T} = 
 2iIm(g^R) \tanh\frac{\epsilon}{2T} = 
-2\pi i  N_{\vp_s}(\vR,\vp_f;\epsilon) \tanh\frac{\epsilon}{2T} 
$. 
}

Surface zero-energy states have delta-function peak with 
amplitude $a \Delta_0$, $N_{abs}(\epsilon) = a \Delta_0 \delta(\epsilon)$, 
and they lower the free energy, 
when shifted from zero by a superflow $p_s v_f \gtrsim T$: 
\begin{align}
\begin{split}
F_{abs}(\vR,\vp_s) - F_{abs}(\vR,0) = -2 N_f a \Delta_0 \frac12 
\left<
2T \ln\left[2\cosh\frac{\vp_s \vv_f }{2T}\right] 
-2T \ln 2 
\right> 
\\
\approx 
- 2 N_f a \Delta_0 \frac12 
\left<
|\vp_s \vv_f| 
\right>
= - N_f \frac{2a}{\pi} \Delta_0 p_s v_f
\end{split}
\end{align}
- the ABS contribution is linear in the superflow $p_s$. 

The Doppler-shifted continuum states, on the other hand, result in increase of the local free energy.
Assuming that the gap edge $|\Delta(\vp_f)|$ is larger than the Doppler shift, 
we can make expansion (neglecting possible pairbreaking and non-linear effects 
from the nodal regions):
\be
N(\vR , \vp_f; \epsilon-\vp_s \vv_f) - N(\vR, \vp_f; \epsilon) 
\approx -\vp_s \vv_f \pder{N}{\epsilon} + \frac12 (\vp_s \vv_f)^2 \pder{^2N}{\epsilon^2} \,,
\ee
and take the energy integral in Eq.~(\ref{eq:FEps}). 
When performing energy integration by parts (two times) one 
can see that 
in this case the important role is played by the 
high-energy parts of the spectrum. 
The linear $\vp_s \vv_f$ term will give zero 
after Fermi surface average, and the second term gives 
quadratic in $p_s$ contribution ($T=0$ limit):
\be
F_{cont}(\vR,\vp_s) - F_{cont}(\vR,0)  
= 2 N_f \left< (\vp_s \vv_f)^2 \right> = N_f v_f^2 p_s^2  \,.
\ee

As a result, near surfaces zero-energy bound states 
create favorable conditions for existence of a superflow. 
The effect of the superflow on local density of states is shown in
\figref{fig:dwall}(left). Quasiclassical calculation of full free energy 
density is presented in \figref{fig:dwall}(right). 
Superflow results in a lower free energy at the surface and higher free energy 
in the bulk. In thin films, completely dominated by the midgap states, 
presence of a current lowers total energy. 

\begin{figure}[t]
\centering\includegraphics[width=2.5in]{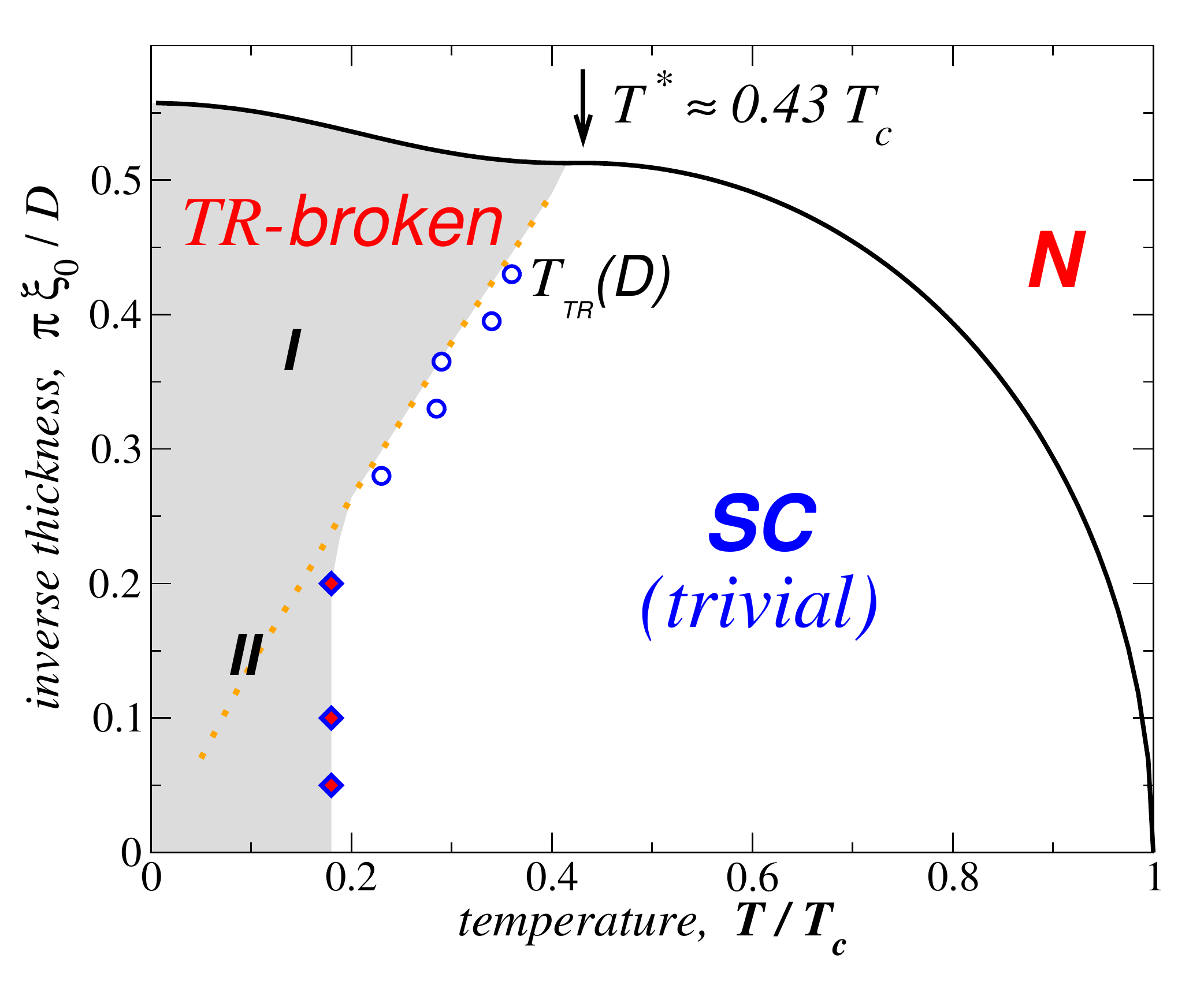}
\centering\includegraphics[width=2.5in]{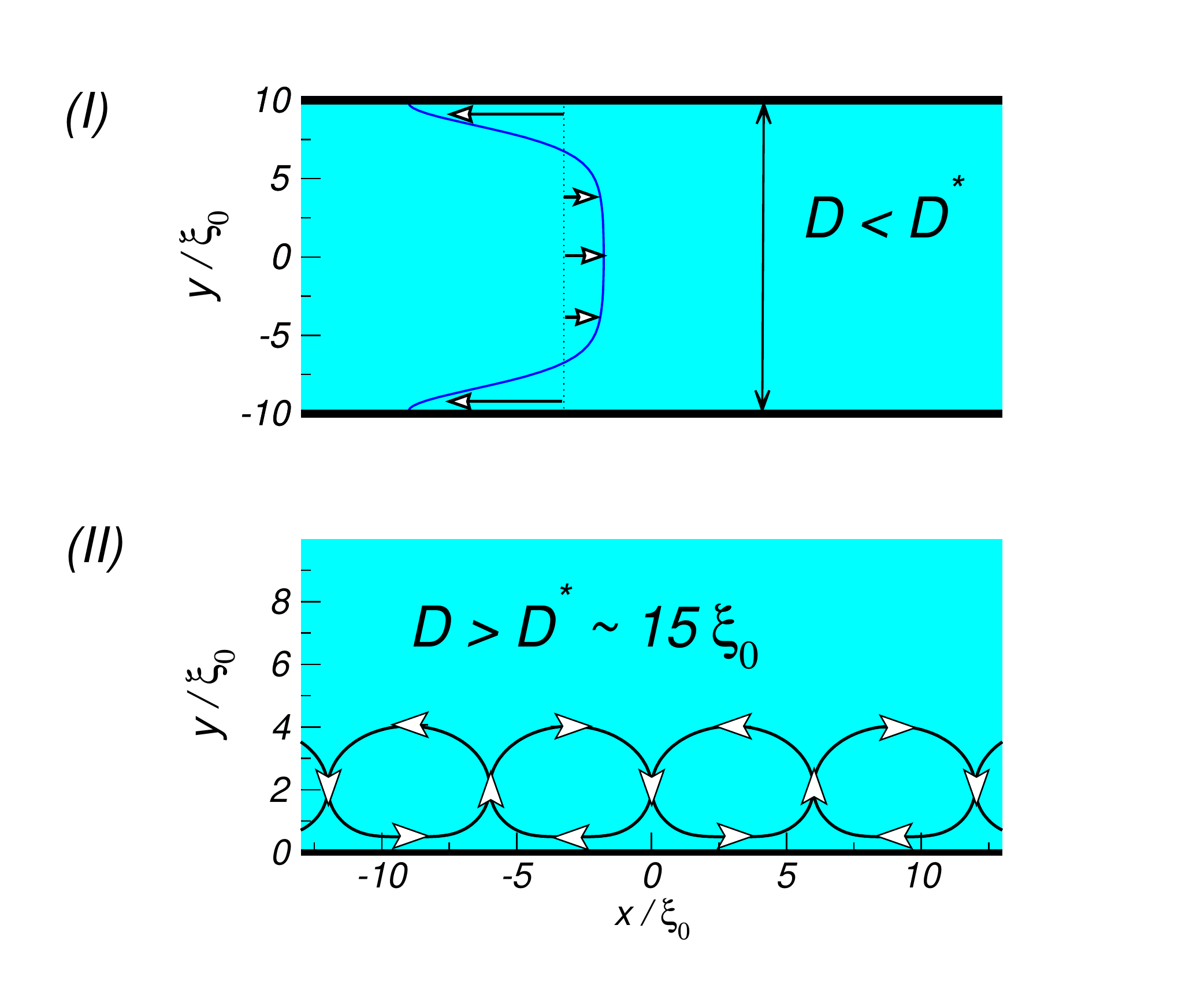}
\caption{
	Phase diagram of a $d$-wave state in film geometry with pairbreaking
	orientation of the surfaces.  In very thin films, surface Andreev
	states spontaneously split producing a flowing current. In films with $D
	\lesssim 15\xi_0$ when states on the two film surface can `feel' each
	other, the current pattern is linear along the film, and onsets at 
	the dotted line, region I \cite{VorontsovAB:2009ef}  
	(open circles are from \cite{Hakansson:2014uf}).
	As the separation between surfaces
	becomes large, the ABS at a given surface re-arrange themselves and
	create a circulating current pattern at each surface, 
	diamonds, region II \cite{Hakansson:2014uf}. 
}
\label{fig:dw}
\end{figure}

Phase diagram of $d$-wave superconductivity in films is shown in \reffig{fig:dw}. The trivial 
superconducting state gives way at low temperature 
to a state with broken time-reversal symmetry. 
The structure of the spontaneously generated currents depends on the size of the 
system. In a small-$D$ sample, non-local interaction of the bound states at different 
edges of the film, lead to currents parallel to the film surfaces. 
The superflow in very thin films is uniform in the cross-section of
the film, $p_s(y)=const$. 
The transition temperature $T_{\scriptscriptstyle TR}(D)$ can be estimated in this regime as 
follows.
The free energy has bound state contribution from the 
boundary regions of size $L_\xi \sim \xi_0$ and the continuum contribution 
from the entire width of the film $D$: 
$F(p_s)-F(0) \approx - N_f \frac{2a}{\pi} \Delta_0 p_s v_f 2 L_\xi 
+ N_f p_s^2 v_f^2 D$. 
The minimum of free energy is given by superflow $p_s \propto L_\xi\Delta_0/v_f D$, and 
the transition temperature into 
into the TR-broken state, 
defined by the splitting of the bound states by this superflow, is 
$T_{\scriptscriptstyle TR} \approx p_s v_f \propto L_\xi \Delta_0 /D$ - linear in $1/D$, 
dotted line in \reffig{fig:dw}(left). This line results in $T_{\scriptscriptstyle TR}=0$ 
for semi-infinite `neutral' superconductor.
In a real, charged superconductors coupling to magnetic field limits the currents 
to the region of magnetic penetration length $\lambda$. The energy balance 
equation becomes 
$F(p_s)-F(0) \approx - N_f (2a/\pi) \Delta_0 p_s L_\xi + N_f p_s^2 v_f^2 \lambda $, 
that results in spontaneous surface currents 
at temperatures below $T_s \propto (\xi_0/\lambda) T_c$ 
even in semi-infinite material \cite{Barash:2000vt}.

However, in samples larger than $10-15 \xi_0$, having a  
uniform, or even exponentially decaying, superflow away from the edges of the film 
becomes energetically too costly, and the 
order parameter adjusts its phase to create a non-uniform two-dimensional pattern 
of circulating currents 
limited only to the surface region \cite{Hakansson:2014uf}. 
The period of the current cell is about the width 
when the bound states on two surfaces `de-couple' from each other, $10-15 \xi_0$.
This indicates that the non-local interactions between bound states is the main driving mechanism 
for TR-broken state, and the characteristic scale for this interaction is $10\xi_0$. 
The temperature that marks onset of current-circulating state becomes 
thickness-independent, as a result, shown by diamonds in phase diagram, \reffig{fig:dw}(left).

\subsection{Multi-component superfluids}
\label{sec:newHe}

In superfluids that have multiple order parameter components, 
structure of the bound states at interfaces is more intricate. 
Various components of the order parameter suppressed differently 
at the boundaries, 
depending on their momentum space basis functions, 
and the non-linear coupling between components can transfer weight between them. 
This interplay creates new ways of adjusting the OP structure 
in confined geometry, and shifting to a new 
non-trivial minima in multi-dimensional energy landscape. 

\begin{figure}[t]
\begin{minipage}{0.4\linewidth}
\centering\includegraphics[width=0.65\linewidth]{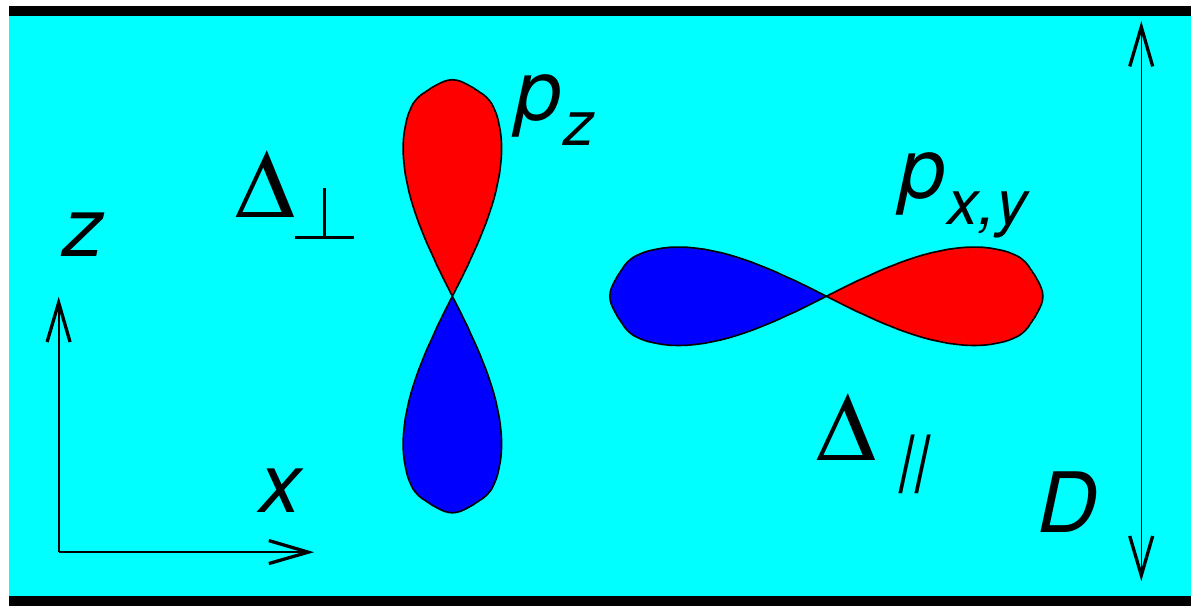}
\centering\includegraphics[width=0.95\linewidth]{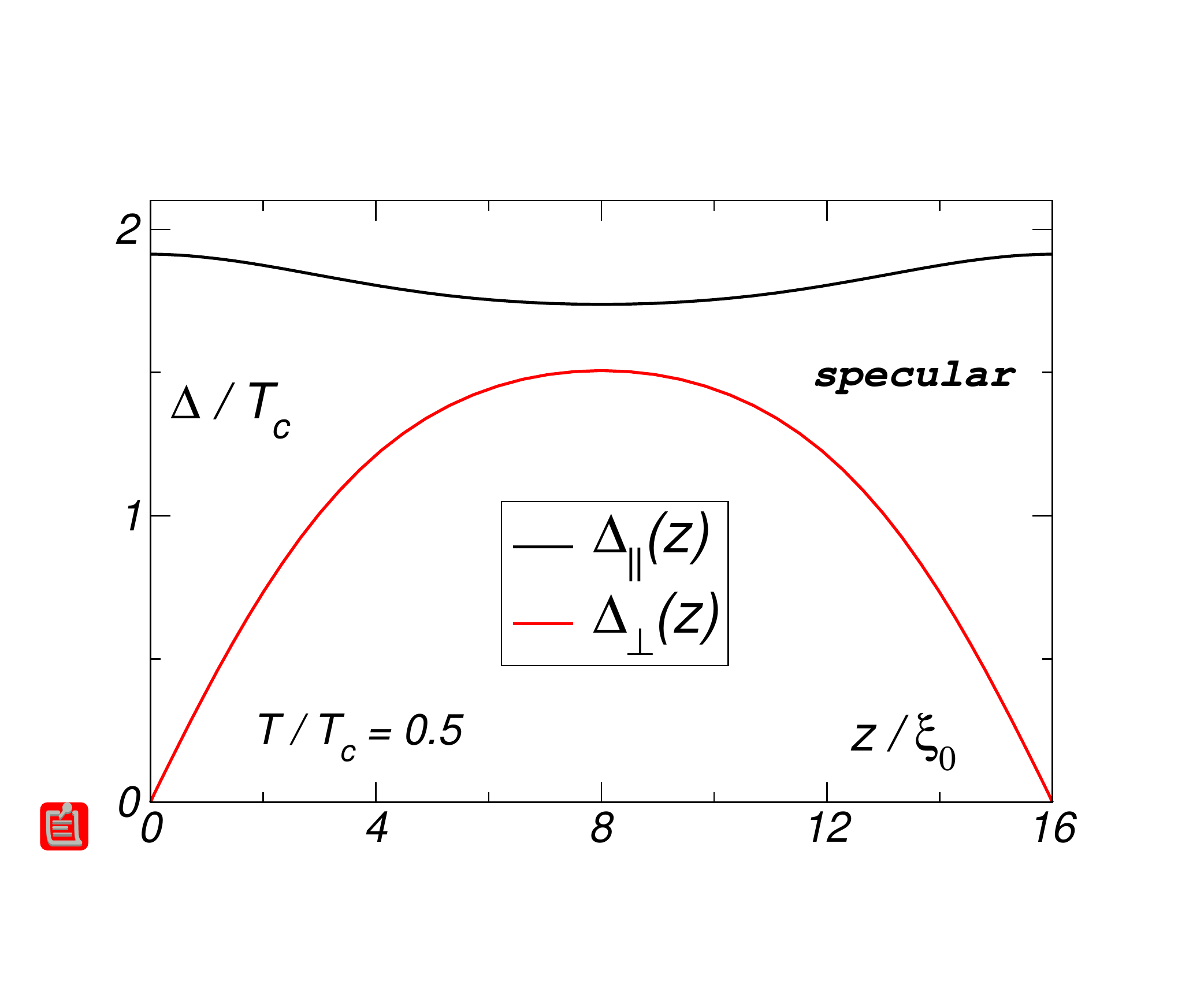}
\end{minipage}
\hfill
\begin{minipage}{0.6\linewidth}
\centering\includegraphics[width=0.98\linewidth]{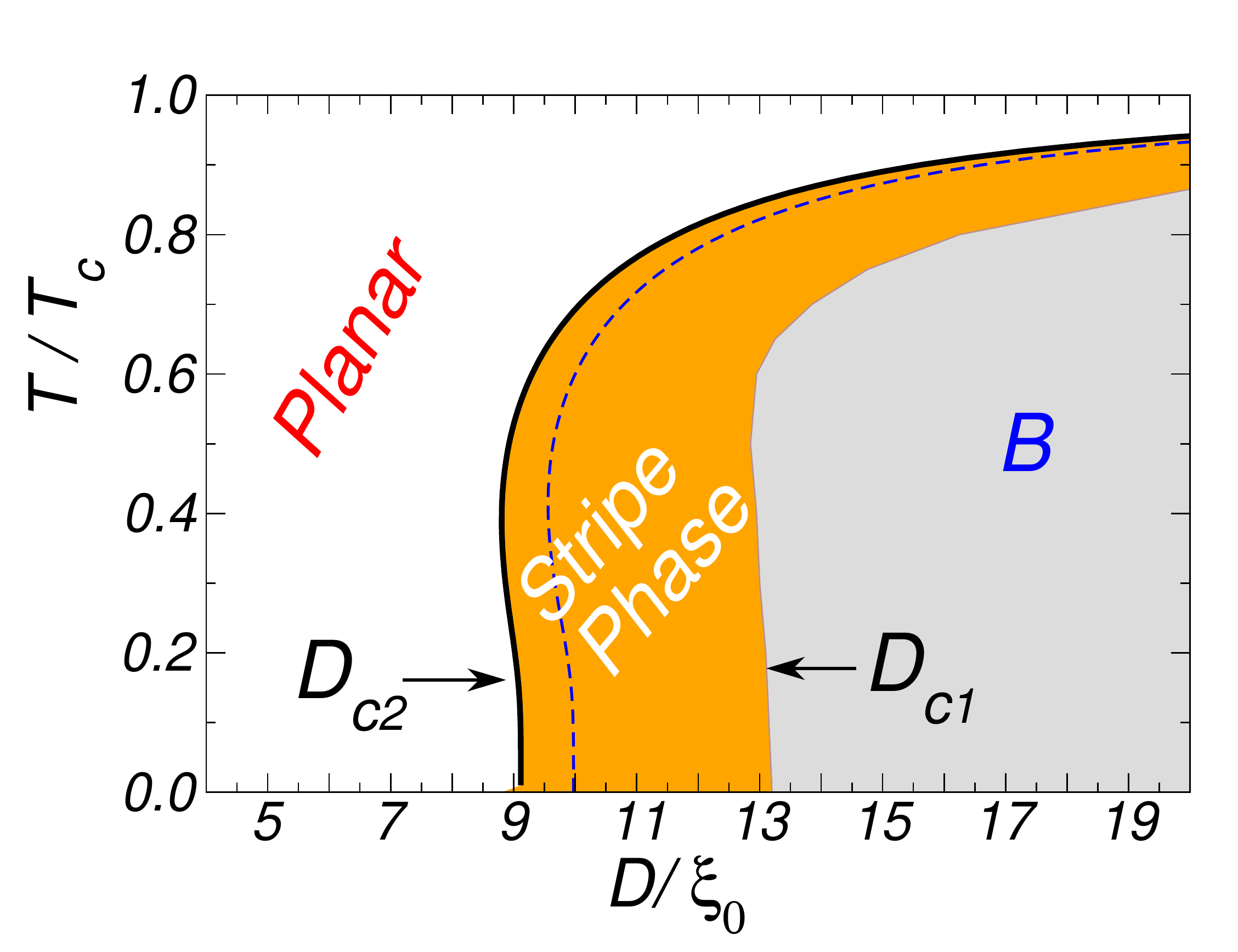}
\end{minipage}
\caption{
	Left: Superfluid \He\ confined to a slab or a film. 
Order parameter component that depends on $\hat p_z$-momentum gets suppressed, 
while `parallel' orbitals result in surface enhancement of corresponding components (specular reflection). 
Right:
transition from the distorted B-phase into Planar phase (weak coupling) 
occurs through a phase that breaks translational invariance along the plane of the film, 
and forms periodic structure with period of several $\xi_0$.\cite{VorontsovAB:2007bs} 
}
\label{fig:dwBfilmPD}
\end{figure}

In spin-1 $p$-wave superfluid \He\ the vector order parameter is parametrized by $3\times 3$ matrix, 
\be
\Delta_\alpha(\vR,\vp_f) = \sum_{i=x,y,z} A_{\alpha i}(\vR) \hat{p}_i \,,
\qquad
A_{\alpha i} = 
\left( \begin{array}{ccc} 
A_{xx} & A_{xy} & A_{xz} \\
A_{yx} & A_{yy} & A_{yz} \\
A_{zx} & A_{zy} & A_{zz} 
\end{array}\right)
\ee
that gives the momentum dependence in terms of orbitals $(\hat{p}_x,\hat{p}_y,\hat{p}_z)$. 
Scattering of the quasiparticles off the $xy$-plane, for example, 
suppresses the orbital-$\hat{p}_z$ OP components $A_{\alpha z}$ 
for specular scattering and 
all components for diffuse scattering. 
Neglecting the dipole-dipole interaction that orients the spin vector relative to the 
orbital vector in a certain way, the stable phase in thick films is the distorted B-phase 
where OP components 
depend only on the transverse coordinate, which traditionally 
is denoted $z$ (rather than $y$ as was in the case of 2D superconductor):
\be
A_{\alpha i}(\vR)_B
=
\left( \begin{array}{ccc} 
\Delta_\parallel(z) & 0 & 0 \\
0 & \Delta_\parallel(z) & 0 \\
0 & 0 & \Delta_\perp(z) 
\end{array}\right)
\ee
The transverse $\Delta_\perp(0)=\Delta_\perp(D)=0$ 
component is pinned by the boundary conditions 
and gets suppressed as film is made thinner, vanishing at 
a second-order transition into Planar phase (weak coupling)
\be
A_{\alpha i}(\vR)_P
=
\left( \begin{array}{ccc} 
\Delta_\parallel^0 & 0 & 0 \\
0 & \Delta_\parallel^0 & 0 \\
0 & 0 & 0
\end{array}\right)
\label{eq:HePop}
\ee
where only $z$-independent parallel components $\Delta_\parallel^0$ remain that do not 
get suppressed by specular scattering. 
This transition is shown by dashed line in \figref{fig:dwBfilmPD}, and also features 
back-bending behavior at low $T$. 

It has been predicted \cite{VorontsovAB:2007bs} that in weak coupling 
in the vicinity of this transition, the superfluid with 
order parameter $A_{\alpha i}(z)$, translationally invariant 
along the film's plane, is unstable towards formation of a new phase that 
spontaneously breaks this symmetry and generates longitudinal modulations 
of the order parameter $A_{\alpha i}(z,x)$,  
forming stripes along some direction in the plane of the film. 

\begin{figure}[t]
\begin{minipage}{0.5\linewidth}
\centering\includegraphics[width=1.1\linewidth]{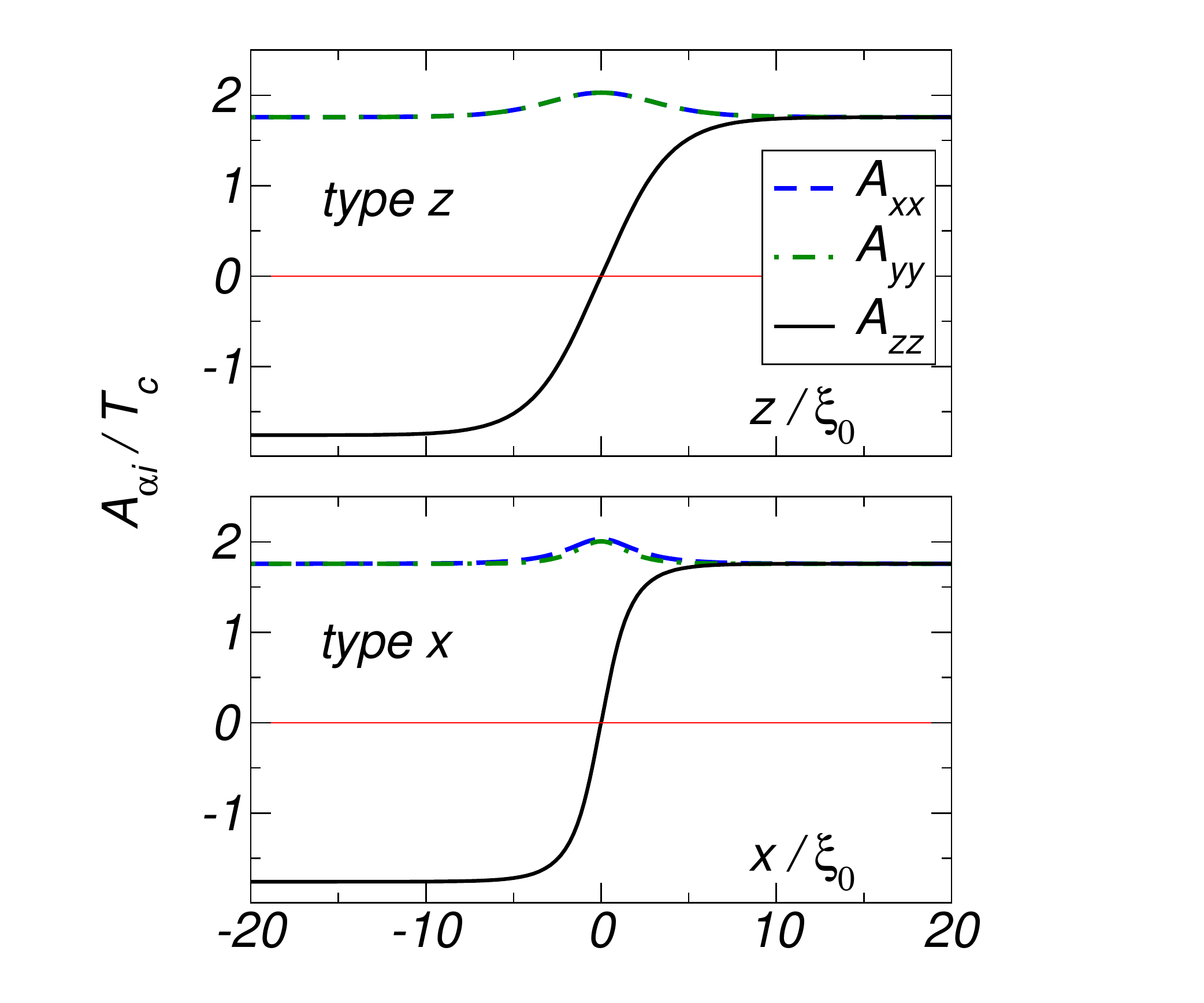}
\end{minipage}
\hfill
\begin{minipage}{0.5\linewidth}
\centering\includegraphics[width=1.1\linewidth]{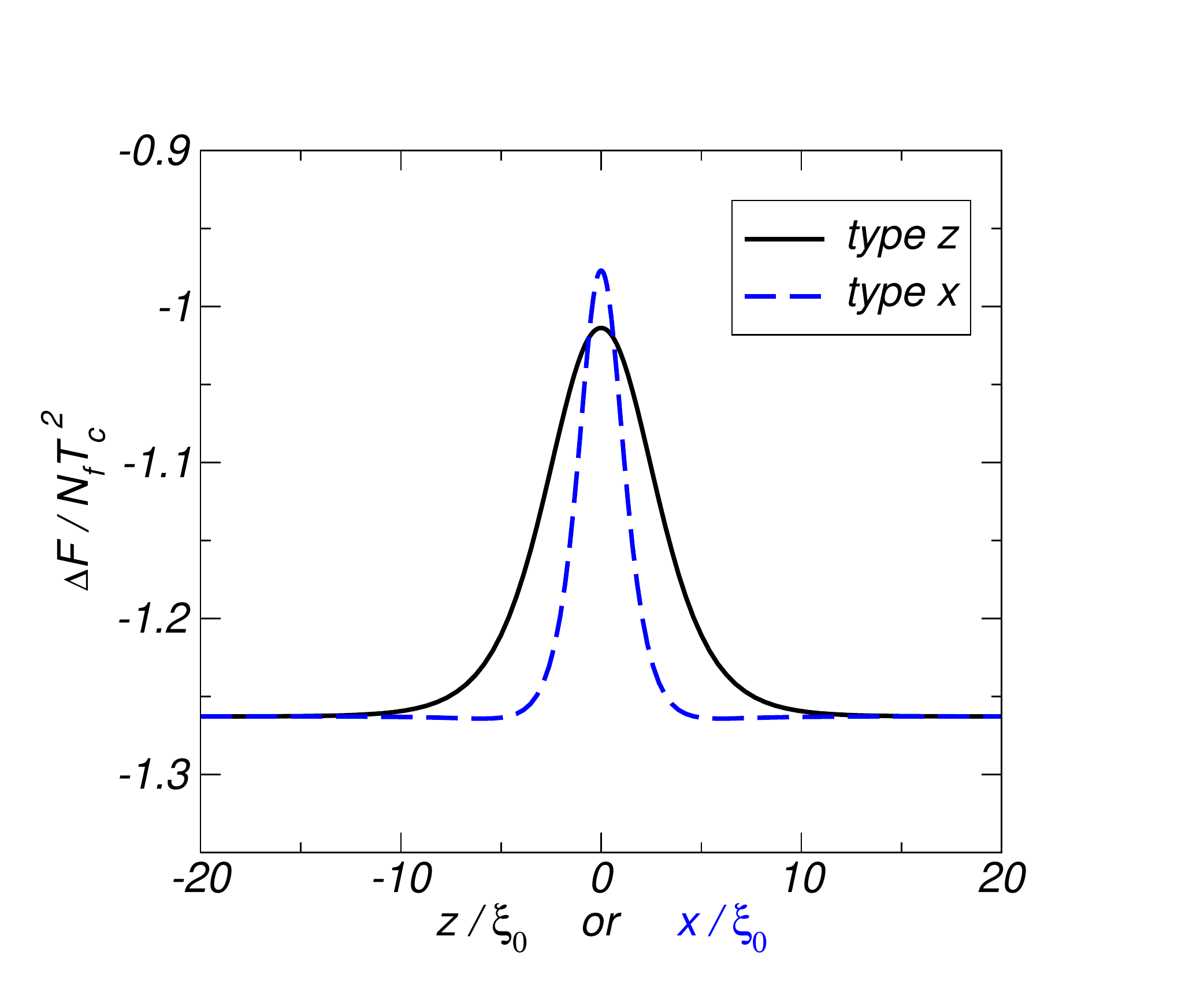}
\end{minipage}
\caption{
	Left: order parameter in domain walls of the B-phase \cite{Vorontsov:2005jy}. 
	These configuration of the OP suppression are not stable in the bulk, 
	but can naturally appear and compete in confined geometry 
	to lower free energy. 
	Right: the free energy profile associated with the domain walls. The longitudinal, 
	type-x,
	wall is narrower and costs less energy. 
}
\label{fig:dwBop}
\end{figure}

The stability and structure of this phase is a consequence of multiple 
inhomogeneous configurations in multi-component superfluids. 
The two relevant configurations for superfluid \He\ are shown in 
\figref{fig:dwBop}. Suppression of the transverse component $A_{zz}(z)=\Delta_\perp(z)$ 
at a pairbreaking surface 
can be mapped onto a domain wall in infinite space where this component 
changes sign $\Delta_\perp(z=\pm\infty) = \pm \Delta_{0B}$. 
This domain wall configuration, which we may call type-z, should be compared with another 
configuration, type-x, when $A_{zz}$ component of the order parameter 
changes along the plane of (a very thick, we can imagine) film. 
The latter configuration costs less in terms of free energy, see  
\figref{fig:dwBop} (right) and is, in fact, the lowest energy domain wall configuration 
\cite{Salomaa:1988da}. 
Although such states are not topologically stable in the bulk \cite{Salomaa:1988da,Silveri:2014wi}, 
pairbreaking geometry creates environment where such configurations can appear  
to minimize the total energy in a finite volume. 
The spectrum of the bound states reflects this interplay, and 
for the two shown domain wall configurations is presented in \figref{fig:dwBdos}. 

\begin{figure}[t]
\centering\includegraphics[width=2.5in]{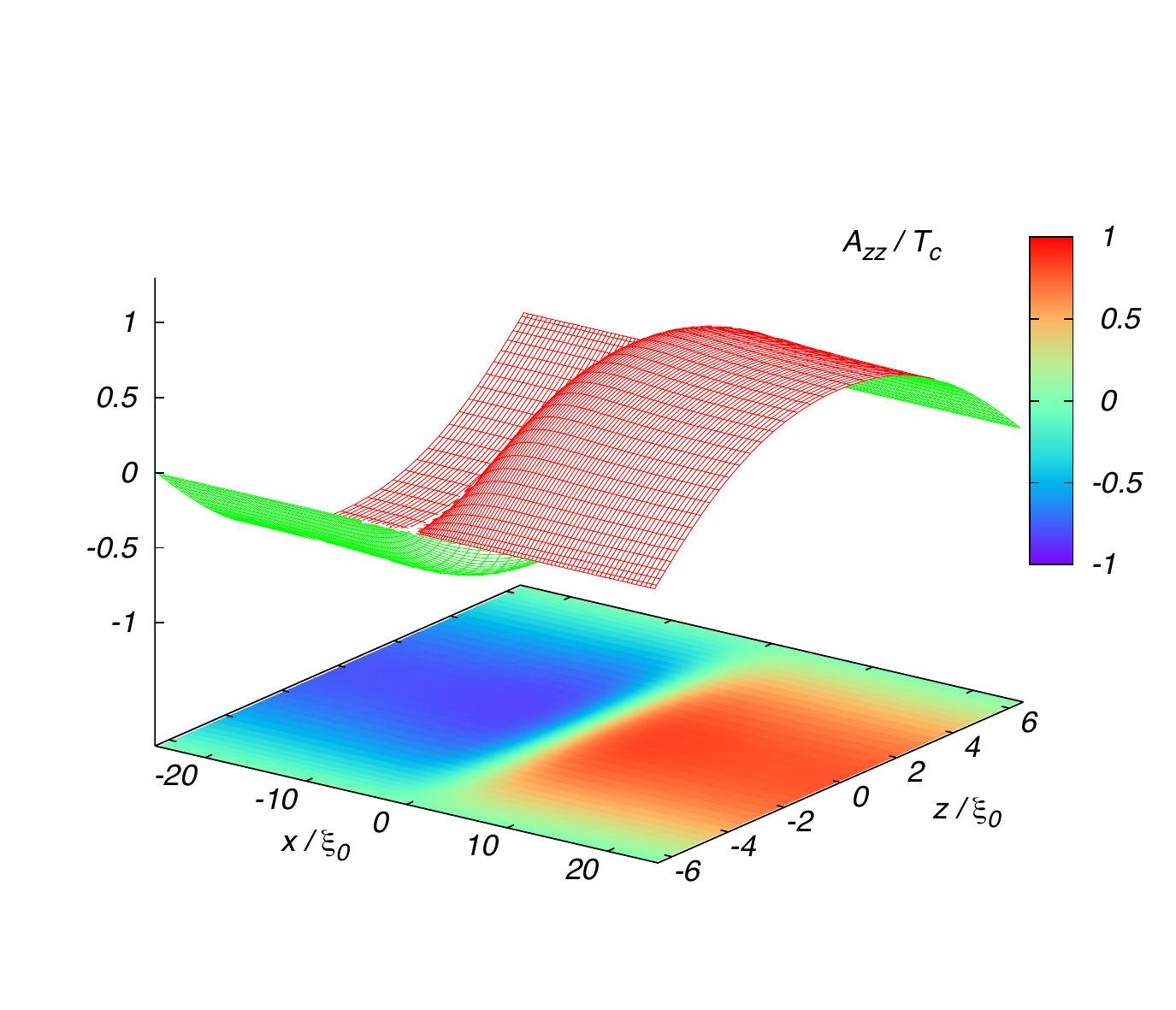}
\centering\includegraphics[width=2.5in]{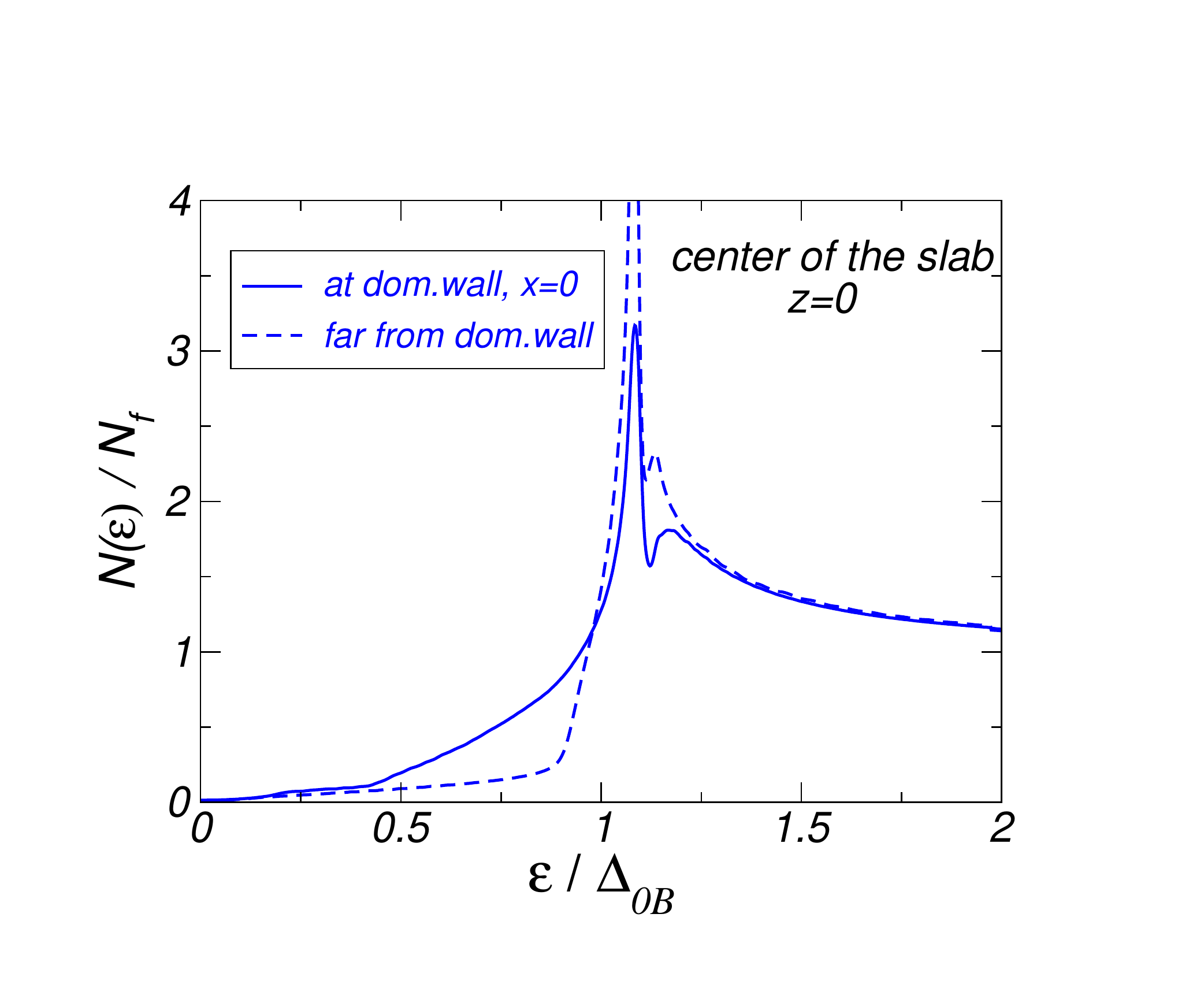}
\centering\includegraphics[width=2.5in]{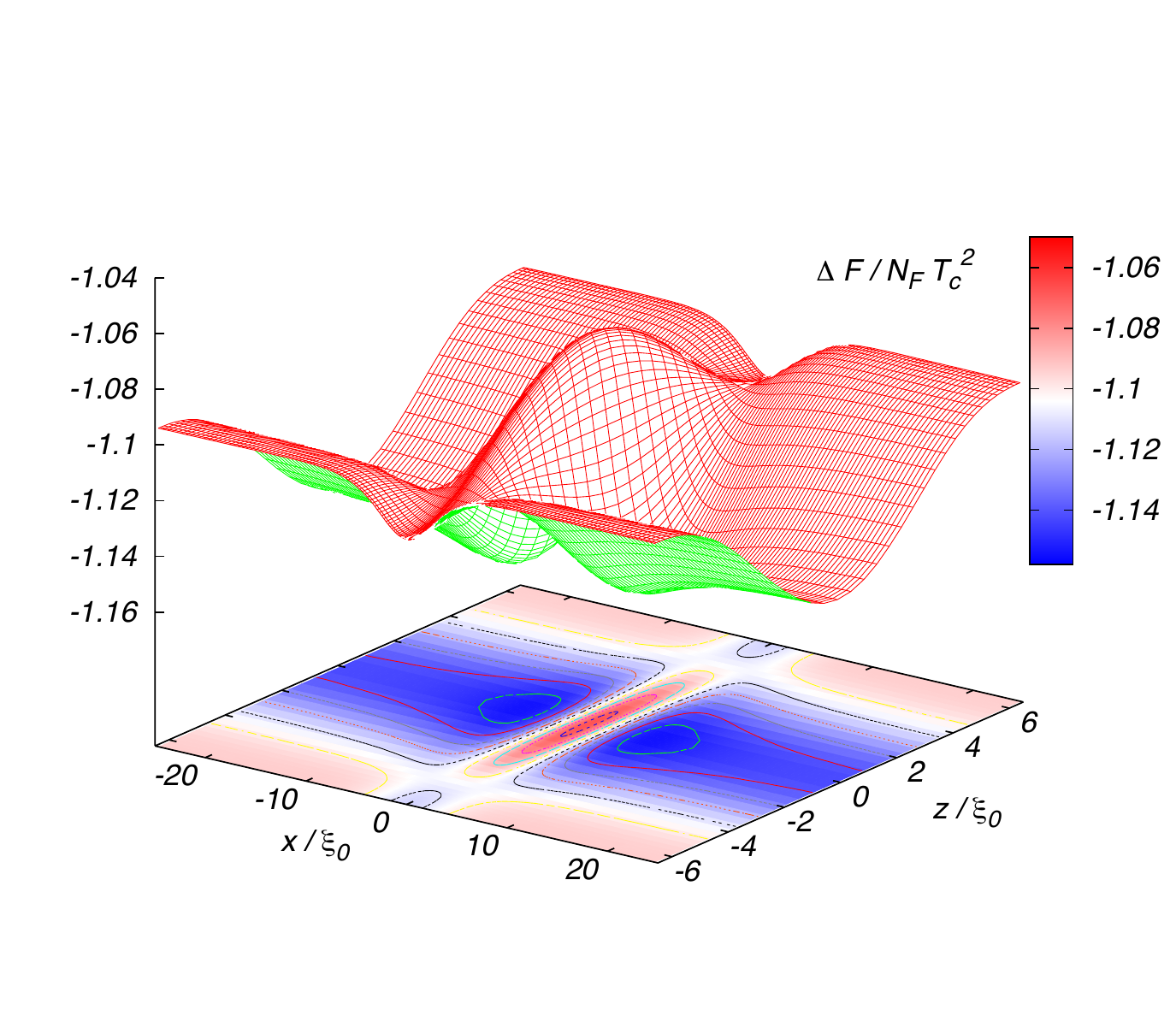}
\centering\includegraphics[width=2.5in]{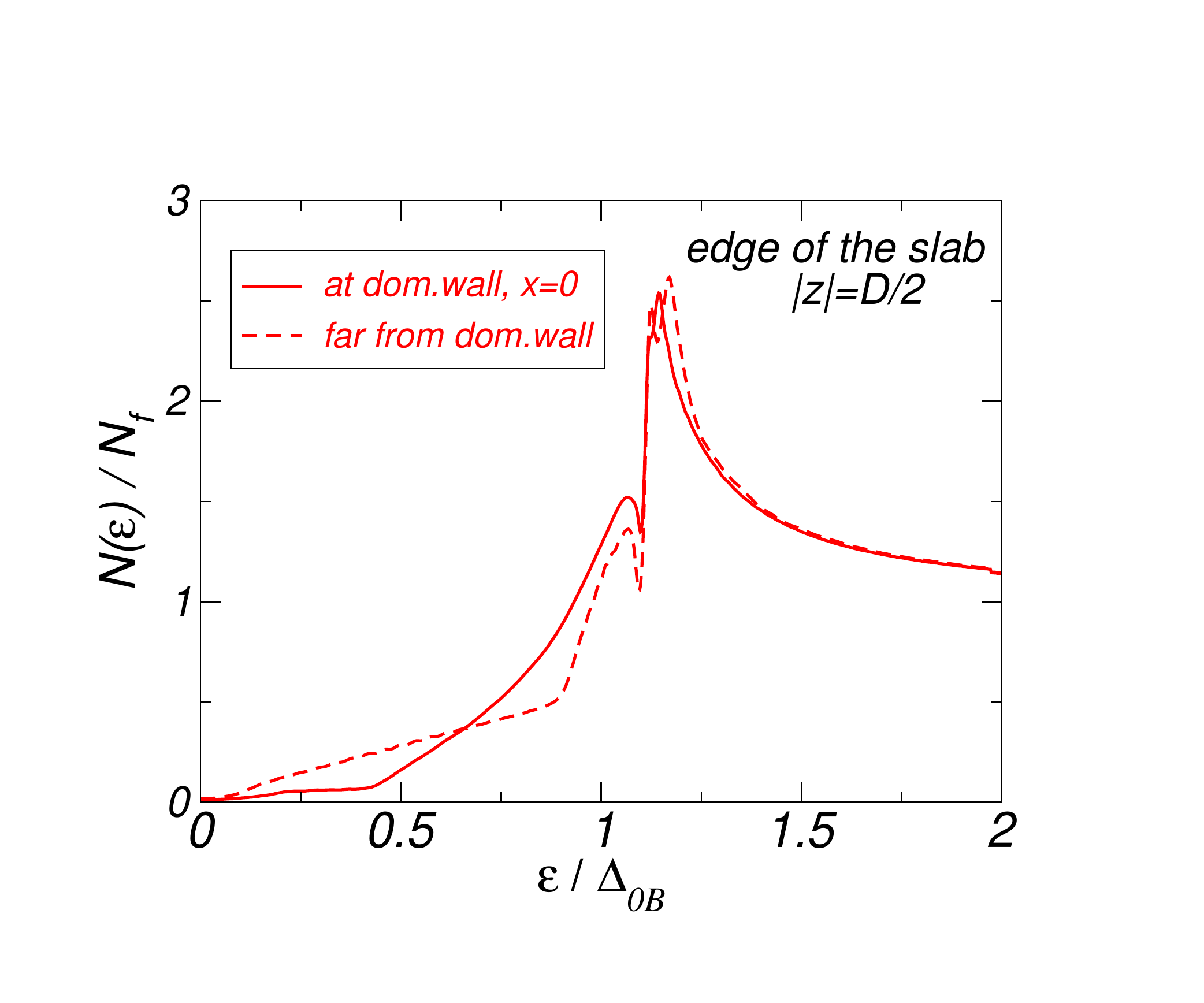}
\caption{
	Top left: pairbreaking associated with the suppression of the 
	$A_{zz}(z) = \Delta_\perp(z)$ component can be reduced by placing additional
	(domain wall) modulation in the film $A_{zz}(x,z)$.  The free energy profile (bottom
	left) shows energy gain at the edges and the center of the film around
	this domain wall. Panels on the right show spectral
	weights far from the domain wall (dashed lines) and at the domain wall
	$x=0$ (solid lines). In the center of the film (top right) the states
	from continuum move to lower energies, and at the edges (bottom right)
	low-energy states shift up, as one approaches domain wall at $x=0$. 
}
\label{fig:dwBfilm}
\end{figure}

In the film geometry, the trivial suppression of the $A_{zz}(z)$ 
across the film is due to reflection $ \hat{p}_z \to -\hat{p}_z$.
This loss of condensation energy, similar to that of type-z domain wall,  
can be reduced by creating additional, type-x, modulation along the film, 
$A_{zz}(z,x)$, shown in top left panel of \figref{fig:dwBfilm}.
This extra modulation `undoes' the trivial pairbreaking by having amplitude sign-change  
$A_{zz}(x>0) \to -A_{zz}(x<0)$ along a trajectory that `bounces' off the film surface near $x=0$. 
The reduction of the pairbreaking appears as energy gains at the T-intersections, 
shown in bottom left panel of \figref{fig:dwBfilm}. 
In sufficiently thin films this energy gain is enough to overcome the extra 
cost of creating the type-x domain wall across the width of the film. 
This energy balance determines the $D_{c1}$ transition in \figref{fig:dwBfilmPD}, where 
a single domain wall enters previously translationally invariant film.
The exact structure of the domain wall is described in \cite{VorontsovAB:2007bs}:
$A_{zz}$ component acquires x-modulation, as we just described, and at the same time a large 
$A_{xz}$ OP component also appears.

In \figref{fig:dwBfilm}, on the right, we can see how the density of states changes 
with the introduction of such a domain wall. Most of the weight redistribution is 
associated with the midgap states. 
As we move into the domain wall region 
along the center of the film, 
states from 
the continuum, $\epsilon \sim \Delta_{0B}$, 
move into the gap region $0.5\Delta_{0B} \lesssim \epsilon \lesssim \Delta_{0B}$. 
On the other hand, near the film edges, the low energy bound states 
shift towards $\Delta_{0B}$. 

To conclude this discussion, we now 
find the instability $D_{c2}$ from the Planar phase into the stripe phase, 
and show that the emergent new order parameter structure at the transition 
agrees with the one that appears at the single domain wall transition $D_{c1}$. 
The second-order $D_{c2}$ instability is defined as the lowest $D$ line where 
new OP structure, different from Planar background phase, emerges. 
We will see that the main difference between this Planar-B transition 
in superfluid \He\ and the N-SC transition in a single-component superconductor, 
is that the existing background condensate \He\ film 
create coupling between the newly generated OP components, 
enforcing certain restrictions on the 
possible structure of the emergent state, even before the fourth order GL terms 
are compared for various phases. 

The procedure to find the instability is to find non-trivial self-consistent solution 
for the order parameter to linear order in perturbation, 
$\hat\Delta_0(\vR,\vp_f) + \hat\Delta_1(\vR,\vp_f)$ 
The unperturbed phase is the Planar state with order parameter 
$\vDelta_0(\vR,\vp_f) = (\Delta_\parallel^0 \hp_x,\; \Delta_\parallel^0 \hp_x,\; 0) $
and the Green's function: 
\be
\hf_0 = 
\pi \frac{\vDelta_0}{\sqrt{\epsilon_m^2 + |\vDelta_0|^2}} (i\vsigma \sigma_y)
\;,\qquad
\hg_0 = 
-\pi\frac{i\epsilon_m }{\sqrt{\epsilon_m^2 + |\vDelta_0|^2}}  \,.
\label{eq:PlanarGF}
\ee
The self-consistency equation for the linear correction to the order parameter  
$\Delta_{1,\alpha}(\vR,\vp_f) = a_{\alpha i}(\vR) \hp_i$ is,
\be
a_{\alpha i}(\vR) \ln {T\over T_c} = T \, \sum_{\epsilon_m} \, 
3 \int \frac{d\Omega_\hvp}{4\pi} \hp_i   
\left( f_{1,\alpha}(\vR,\vp_f; \epsilon_m) - \frac{a_{\alpha j}(\vR) \hp_j }{|\epsilon_m|} \right) \,.
\ee
Linear order correction $f_{1,\alpha}$ to the off-diagonal propagator is found from 
Eilenberger equations and normalization conditions, 
for $\hat g$ and $\hat f$ components.
Breaking the commutator in (\ref{eq:eilenberger}) into components we have
\begin{align}\label{eq:eilbycomp}
\begin{split}
i\vv_f\cdot\grad \hg + (-\hDelta \underline{\hf} + \hf \underline{\hDelta}) = 0
\, , \qquad 
\hg^2 + \hf \underline\hf = -\pi^2 \,,
\\
i\vv_f\cdot\grad \hf + 2i\epsilon_m \hf + ( - \hDelta \underline{\hg} + \hg \hDelta) = 0
\, , \qquad 
\hg\hf + \hf \underline\hg = 0 \,.
\end{split}
\end{align}
We linearize them with respect to $(\hg_1, \hf_1, \hDelta_1)$,
and in the normalization conditions we use expressions for 
the background Green's functions (\ref{eq:PlanarGF}) together with 
symmetry relations (\ref{eq:tau1symm}): 
\begin{align}
\begin{split}
& (2i\epsilon_m + i\vv_f\cdot\grad ) \hf_1 +  (- \hDelta_0 \underline{\hg}_1 + \hg_1 \hDelta_0) = 
\hDelta_1 \underline\hg_0 - \hg_0 \hDelta_1 
\;, \qquad 
\hg_1 \hDelta_0 + \hDelta_0 \underline{\hg}_1 = 0
\\
& i\vv_f\cdot\grad \hg_1 + (-\hDelta_0 \underline{\hf}_1 + \hf_1 \underline{\hDelta}_0) = 
(\hDelta_1 \underline{\hf}_0 - \hf_0 \underline{\hDelta}_1) 
\;, \qquad 
-2i\epsilon_m \hg_1 + \hf_1 \underline\Delta_0 + \hDelta_0 \underline{\hf}_1 = 0
\end{split}
\end{align}
Here we treat the gradient terms non-perturbatively, 
assuming they have same magnitude as the other energy terms. 
Combining pairs of equation in each line we eliminate 
$\underline{\hat g}_1$ and $\underline{\hat f}_1$:
\begin{align}
\begin{split}
& \left( i\epsilon_m + \frac{i}{2} \vv_f\cdot\grad \right) \hf_1 +  \hg_1 \hDelta_0 =  - \hg_0 \hDelta_1 
\,, \\
& \left(-i\epsilon_m + \frac{i}{2} \vv_f\cdot\grad \right) \hg_1 + \hf_1 \underline{\hDelta}_0 = 
\frac12 (\hDelta_1 \underline{\hf}_0 - \hf_0 \underline{\hDelta}_1) 
\,.
\end{split}
\end{align}
Using the fact the unperturbed solution is uniform, $\grad_\vR \widehat g_0=0$, we can use 
the second equation to eliminate $\hg_1$ from the first, 
\begin{align}
\begin{split}
\left(-i\epsilon_m + \frac{i}{2} \vv_f\cdot\grad \right) 
\left( i\epsilon_m + \frac{i}{2} \vv_f\cdot\grad \right) \hf_1 
+ \left( -\hf_1 \underline{\hDelta}_0 
+ \frac{\hDelta_1 \underline{\hf}_0 - \hf_0 \underline{\hDelta}_1}{2} \right) \hDelta_0 
\\
=  - \hg_0 \left(-i\epsilon_m + \frac{i}{2} \vv_f\cdot\grad \right) \hDelta_1 
\end{split}
\end{align}
Substituting unitarity condition 
$\underline\hDelta_0 \hDelta_0 = - |\vDelta_0|^2$, 
and expressions for $\hg_0, \hf_0$, one derives final equation for corrections to 
propagator, linear in order parameter deviations, 
\begin{align}
\begin{split}
\left[ \epsilon_m^2 + |\vDelta_0|^2 + \left(\frac{i}{2} \vv_f\cdot\grad\right)^2 \right] \hf_1 
& = \frac{\pi}{\sqrt{ \epsilon_m^2 + |\vDelta_0|^2 }} 
 \\ 
& \hspace*{-2cm} \times
\left[ 
i \epsilon_m \left(-i\epsilon_m + \frac{i}{2} \vv_f\cdot\grad \right) \hDelta_1 
 + \frac12 \left(  |\vDelta_0|^2 \hDelta_1 + \hDelta_0 \underline\hDelta_1 \hDelta_0 \right)
\right]
\,.
\end{split}
\end{align}
The gradient term on the right-hand side acting on the order parameter can be dropped 
since it is odd in energy and will disappear after $\pm\epsilon_m$ energy summation in the 
self-consistency equation. 
Equation for the off-diagonal vector $\vf_1(\vR, \vp_f; \epsilon_m)$ 
in terms of the order parameter $\vDelta_1(\vR,\vp_f)$, 
\be
\left[ \epsilon_m^2 + |\vDelta_0|^2 + \left(\frac{i}{2} \vv_f\cdot\grad\right)^2 \right] \vf_1 
 = \pi \frac{
\left[ \epsilon_m^2 + \frac12  |\vDelta_0|^2 \right] \vDelta_1  - \vDelta_0 (\vDelta_0 \vDelta_1^*) + \frac12 (\vDelta_0^2)  \vDelta_1^* 
	 }{\sqrt{ \epsilon_m^2 + |\vDelta_0|^2 }} 
\ee
couples, through finite $\vDelta_0$, solutions for $\vDelta_1$ and its 
complex conjugate $\vDelta_1^*$. 
As a result, spatially dependent $e^{\pm i\vq\vR}$ solutions exist in pairs 
and a current-carrying solution is not possible. 
We use Fourier expansion to solve for $\vf_1$ in terms of OP harmonics: 
\be
a_{\alpha i}(\vR) = \sum_\vq a_{\alpha i}(\vq) e^{i\vq \vR}  
\qquad \Rightarrow\qquad
f_{1,\alpha}(\vR,\vp_f;\epsilon_m) = \sum_\vq f_{1,\alpha}(\vq,\vp_f;\epsilon_m) e^{i\vq \vR}  
\ee
and write coupled self-consistency equations for 
the pairs $a_{\alpha i}(\vq)$ and $a_{\alpha i}(-\vq)^*$, 
\be
\ln \frac{T}{T_c} 
\left( \begin{array}{c} a_{\alpha i}(\vq) \\  a_{\alpha i}(-\vq)^*  \end{array} \right) 
=
\left( \begin{array}{cc} 
K_{ij}^{\alpha \beta}(\vq,\vq) &  K_{ij}^{\alpha \beta}(\vq,-\vq) \\ 
K_{ij}^{\alpha \beta}(\vq,-\vq)^* &  K_{ij}^{\alpha \beta}(\vq,\vq) 
 \end{array} \right) 
\left( \begin{array}{c} a_{\beta j}(\vq) \\  a_{\beta j}(-\vq)^*  \end{array} \right) 
\label{eq:PlanarInstab}
\ee
where the $a$-columns consist of 18 complex numbers, 
and the $K$-matrix is $18 \times 18$.  Its elements are, 
\begin{align}
\begin{split}
& K_{ij}^{\alpha \beta}(\vq,\vq) = 2\pi T \, \sum_{\epsilon_m>0} \, 
3 \int \frac{d\Omega_{\vp_f}}{4\pi} \hp_i \hp_j  
\left[ \left( \frac{E_m}{E^2_m(\vq)} - \frac{1}{\epsilon_m}\right) 
- \frac{|\vDelta_0(\vp_f)|^2}{2 E_m E^2_m(\vq)}
\right] \delta_{\alpha\beta} \,,
\\
& K_{ij}^{\alpha \beta}(\vq,-\vq) = 2\pi T \, \sum_{\epsilon_m>0} \, 
3 \int \frac{d\Omega_{\vp_f}}{4\pi} \hp_i \hp_j  \frac{1}{ E_m E^2_m(\vq)}
\left[  \frac12 \vDelta_0(\vp_f)^2 \delta_{\alpha\beta}  - \vDelta_0(\vp_f)_\alpha \vDelta_0(\vp_f)_\beta
\right]\,,
\end{split}
\end{align}
where shorthand 
$E_m = \sqrt{\epsilon_m^2 + |\vDelta_0(\vp_f)|^2 }$ 
and
$E_m(\vq) = \sqrt{\epsilon_m^2 + |\vDelta_0(\vp_f)|^2 + \left(\frac12 \vv_f\vq\right)^2 }$ 
are used. 
The $K$-matrix is Hermitian and we numerically find all non-trivial solutions to  
Eq.~(\ref{eq:PlanarInstab}).
At a given $T$ we vary $q_x$ to find maximal $q_z$ 
that satisfies $|\hat K(T,q_x,q_z) - \hat 1 \ln(T/T_c)|=0$, 
and find the eigenvector $(a_{\alpha i}(\vq), a_{\alpha i}(-\vq)^*)$ for the pairs $(q_x,q_z)$.
We will then impose a boundary condition that all components with $\hat{p}_z$ orbital dependence 
vanish at the specular edges of the slab for any $x$:
\be
a_{\alpha z}(x, z=\pm D/2) = \sum_{q_z, q_x} a_{\alpha z}(\vq) e^{i (q_z D/2 + q_x x) } =0 \,. 
\ee
This procedure determines 
the transition values of the wave vector, which we denote by 
capital letters $Q_z(T)$ and $Q_x(T)$, and
the structure of the nucleating order parameter at the transition.
For transition from the Planar phase,
we find two degenerate solutions (due to z-reflection symmetry) 
given by pair 
$(a_{\alpha i}(\vQ_1), a_{\alpha i}(-\vQ_1)) = \{-a_{zx}, a_{zz} \}$ with 
$\vQ_1 = (Q_x, Q_z)$,
and pair
$(a_{\alpha i}(\vQ_2), a_{\alpha i}(-\vQ_2)) = \{a_{zx}, a_{zz} \}$ 
with $\vQ_2 = (Q_x, -Q_z)$. 
The (real) components of the eigenvectors have amplitudes
$a_{zz} =1, \; a_{zx} \approx 0.6$ in the temperature range $0.2 < T/T_c < 0.6$. 
The two are different by the relative sign between the $a_{zx}$ 
and $a_{zz}$ components, as a result of $Q_z$ inversion. 
These two solutions describe 
transition from the Planar state into a new state: 
\be
A^{(0)}_{\alpha i} = \Delta_0
\left( \begin{array}{ccc} 
1 & 0 & 0 \\
0 & 1 & 0 \\
0 & 0 & 0 
\end{array}\right) 
\qquad \Longrightarrow \qquad  
A^{(0)}_{\alpha i} + 
a_{\alpha i}(\vR)_{\vQ_1} + a_{\alpha i}(\vR)_{\vQ_2} 
\ee
with the structure of the nucleated order parameter, 
\begin{eqnarray}
a_{\alpha i}(\vR)_{\vQ_1} = \Delta_1 
\left( \begin{array}{ccc} 
0 & 0 & 0 \\
0 & 0 & 0 \\
-a_{zx} & 0 & a_{zz} 
\end{array}\right) 
\left( e^{i (Q_z z + Q_x x)} +  e^{-i (Q_z z + Q_x x)} \right) \,,
\\
a_{\alpha i}(\vR)_{\vQ_2} = \Delta_2 
\left( \begin{array}{ccc} 
0 & 0 & 0 \\
0 & 0 & 0 \\
a_{zx} & 0 & a_{zz} 
\end{array}\right) 
\left( e^{i (-Q_z z + Q_x x)} +  e^{-i (-Q_z z + Q_x x)} \right) \,. 
\end{eqnarray}
To satisfy the boundary condition $a_{zz}(x,z=\pm D)$ 
the total solution must be a combination of these two, 
with equal amplitudes $\Delta_1 = \Delta_2$, so that 
\begin{eqnarray}
a_{\alpha i}(\vR) = a_{\alpha i}(\vR)_{\vQ_1} + a_{\alpha i}(\vR)_{\vQ_2} 
= \Delta'
\left( \begin{array}{ccc} 
0 & 0 & 0 \\
0 & 0 & 0 \\
a_{zx} \sin Q_z z \sin Q_x x& 0 & a_{zz} \cos Q_z z \cos Q_x x
\end{array}\right) 
\end{eqnarray}
and the boundary condition $\cos Q_z D/2=0$ 
giving the smallest film thickness for maximal $Q_z$, 
\be
D = \frac{\pi}{Q_z} \,.
\ee
This structure agrees with the order parameter obtained numerically in \cite{VorontsovAB:2007bs}.
Note, that any other eigenmodes would have lower $Q_z$ 
and would give transition in thicker films; also the 
opposite choice for the relative overall sign between 
$\vQ_1$,$\vQ_2$ solutions, $\Delta_1 = -\Delta_2$, 
would give $A_{zz} \sim \pm \sin Q_z D/2=0$, which would again lead to 
a thicker film, $D = 2\pi/Q_z$. 

\subsection{Other geometries, pairing states, and effects of competing interactions 
and boundary conditions} 
\label{sec:newO}

From the two cases above, we see that the structure and stability of new phases 
in confined geometry depends on the symmetry and number of OP components.
Any interaction that creates additional energy shifts of the midgap states, modify their weights, 
and affects OP suppression, 
will have a significant effect on the free energy landscape and appearance of new phases. 

For example in the $d$-wave films, 
the structure of the non-uniform states that is generated in the films is determined 
by the $1/(i\epsilon_m-\vq\vv_f/2)$ factors in Eqs.~(\ref{eq:fe}), 
that can be interpreted as Doppler-shifted 
quasiparticles' energy due to non-uniform order parameter. 
This shift will be strongly affected by any changes in the Fermi surface shape through 
direction of Fermi velocity $\vv_f$ and the anisotropy of the density of states that 
appear in the FS angle integrals. Work \cite{Miyawaki:2015gg} demonstrated that 
if the Fermi surface has square shape with flat nested regions and the 
$d$-wave nodes residing on its sharp ends, the broken time-reversal state 
is stabilized in a much thinner films. 

Similar energy shifts occur when there is applied magnetic field. In a case of 
Zeeman interaction between external field $H$ and electron spin moment, 
the shifts are isotropic in momentum space 
$1/(i\epsilon_m-\vq\vv_f/2 \pm \mu_B H)$, which creates interplay between 
non-uniform effects and Pauli-breaking effects, resulting in rich phase diagram that 
features re-entrant superconductivity \cite{Hachiya:2013en}. 

Random scattering effects, either by impurities, and in particular by the 
atomically rough surfaces, are known to deform the spectrum of midgap states 
in a significant way. 
Calculations have shown, however, that 
the spontaneous current-carrying state is relatively robust toward scattering.
For example, to completely suppress this state, 
impurity concentrations that give mean free path $\ell/\xi_0 \sim 5$ and 
60\% $T_c$ suppression, are needed \cite{VorontsovAB:2009ef}. 
A calculation \cite{Higashitani:2015ic} with continuously adjusted surface specularity parameter 
$S$ has shown that the current-carrying state is suppressed when 
specular reflection probability is $S \lesssim 0.2$ - very close to diffuse limit $S=0$. 
Re-orientation of the crystal axes 
relative to the surface plane, $\cY(\phi) = \sin2(\phi-\phi_0)$, reduces 
phase space for trajectories that produce zero-energy states. 
To eliminate spontaneous currents misalignment 
angles $\phi_0 \sim 23^\circ$, that reduce the weight of zero-energy states in about half, 
are required \cite{Hakansson:2014uf}. 

In multi-component systems, a phase with broken continuous translations 
has been described in \He-B, confined to narrow cylinders. \cite{Aoyama:2014uc}
Recent work by Northwestern University theory group, 
has shown that 
in Ginzburg-Landau regime the strong-coupling effects in \He\ 
do not eliminate the stripe state in film geometry. 
They also have discovered that a similar instability 
towards formation of periodic order in two-dimensional strips 
of chiral $p$-wave state $\Delta_x \hat{p}_x + i \Delta_y \hat{p}_y$.

\section{Experimental investigation of thin films}
\label{sec:experim}

It has been realized very early that Andreev bound states and spontaneous surface currents 
can help determine symmetry and structure of the order parameter in new superconductors. 
Their properties have been 
studied and used in high-T$_c$ materials, in particular in tunneling experiments 
\cite{Lofwander:2001uk,Deutscher:2005tm}. 
Most of the work concentrated on description of 
subdominant pairing channels, as a way of identifying the symmetry and origin of the 
pairing state in cuprates, since unusual surface properties can give clear signature 
of such pairing states.\cite{Carmi:2000ct,Elhalel:2007tj,Saadaoui:2011we,Gustafsson:2012ii} 

In superfluid \He, that has one of the most complex order parameter 
structures out of all known superconductors, investigation of 
Andreev surface states is particularly interesting. 
Despite extensive work, 
up to date there have been few experiments directly testing their presence. 
Measurement of transverse impedance in a series of experiment 
\cite{Aoki:2005ub,Saitoh:2009uj,Okuda:2012wg}
demonstrated importance of 
surface states for mechanical coupling between the 
transducers and the oscillations of the liquid, 
and determined their signatures for varying surface roughness. 
Specific heat measurements of superfluid \He\ in silver heat exchanger constructed 
of sintered silver particles 
looked at thermodynamic properties of the Andreev states. 
\cite{Choi:2006jr}

Small scale devices, where pairbreaking and bound states dominate physical properties, 
provide a unique aspect for investigation of unconventional superfluid 
and superconducting phases, and can provide valuable insight into properties 
of the surface bound states themselves. 
While film geometry in superconductors can be considered unusual, 
superfluid \He\ provides a more natural ground for investigation of complex condensates 
in restricted geometries, such as films or slabs. 
For this reason experimental 
investigation of 
superfluid \He\ phases in slabs or films has started early, following 
theoretical development in 1970-80s. 

After detection in 1985 of superfluidity in thin \He\ films 
formed on vertical walls \cite{Sachrajda:1985uf}, 
a number of different techniques was applied to study their properties. 
A film formed on an elevated substrate \cite{Davis88} was used to measure 
suppression of the superfluid transition temperature, and flow properties. .  
Critical current, superfluid density were measured by flow over beaker rim 
\cite{DAUNT:1988vn,Steel:1994dr}. 
Torsional oscillators \cite{XuCrooker} were used to determine $T_c(D)$ dependence, 
and superfluid density. 
An innovative way to excite and detect 
third sound waves in superfluid films was used in \cite{Schechter:1998ub}. 
Another dynamical excitation method by inter-digitated capacitors 
was employed to study flow in films\cite{Saitoh:2003kz,Saitoh:2004tl,Saitoh:2007gp}. 

Phase transitions between \He\ phases in confinement, driven by 
OP suppression, is typically investigated using 
pressure dependence of the coherence length $\xi_0(p)$ that 
changes from 77 nm at zero pressure to 15 nm at the melting pressure, 
making it possible for manipulation of effective film thickness $D/\xi_0(p)$ through 
pressure variation. 
NMR measurements in stacks of Mylar sheets reported measurements of 
superfluid density and identified the thin-film phase to be 
the A-phase in 0.3 $\mu$m slabs \cite{Freeman:1988}. 
Measurements \cite{Kawae:1998wx} reported A-B transition in a
stack of slabs with distribution of thickness from around 0 to $\sim 1.5\mu$m.
The pressure dependence of A-B transition was also investigated in \cite{Miyawaki:2000vo} 
using more uniformly separated polyethylene films of $1.1\pm0.3\mu$m. 
In a similar experiment the A-B transition was mapped in $0.8\pm0.04\mu$m as a 
function of pressure \cite{Kawasaki:2004kp}.

The last decade saw several significant advances on the experimental side. 
The techniques were developed for more precise nano-fabrication 
that promise a better look at the underlying physics of confined superfluids. 
Several different types of 
experimental cells have been developed with the goal of studying 
of superfluid in a single slab. 
Single-film devices allow for better control of uniform thickness.  
Due to precisely defined geometry and dimensions 
single-film cells have better ability to investigate ABS, 
and they can also be manufactured with purpose of 
dynamical excitation of surface ABS.
Royal Holloway University of London group pioneered new ways to fabricate  
single-slab nanofluidic cavity and to perform high-sensitivity NMR measurements 
in a small-volume systems and measured $T_c$ suppression and A-B transition 
\cite{Levitin:2013hq,Levitin:2014jd}.
The team at University of Florida built 
micro-electro-mechanical systems (MEMS) oscillators that can dynamically excite 
quasiparticles in thin films 
\cite{Gonzalez:2011ez,Gonzalez:2013ez,Gonzalez:2013ji}.
The low-temperature group at University of Alberta produced 
and started testing nano-mechanical resonator \cite{Rojas:2015fm}. 
These techniques are well-suited to explore 
whether there is evidence of new unusual phases, 
and to advance our understanding of properties of the surface states and 
superfluid phases in confinement, 
where several discrepancies between theory and experiment 
persist and need to be resolved. 

Suppression of the critical temperature of superfluid, $T_c(D)$, due to 
diffuse surface scattering is potentially one of these anomalies. 
The earlier flow \cite{DAUNT:1988vn,Steel:1994dr} 
and torsion oscillator \cite{XuCrooker} measurements had large error bars that 
agreed reasonably well with theoretical prediction of suppression \cite{KJALDMAN:1978vo}. 
A more recent data of two different groups \cite{Saitoh:2003kz,Levitin:2013hq} 
show noticeable deviations from theoretical predictions and 
surprisingly larger than expected suppression 
in both low and high temperature regimes. 
A more detailed description of some of the experiments is summarized in Fig.~10 of \cite{Levitin:2014jd}. 

There is also no consistency between various theoretical and experimental conclusions 
about the confinement-induced A-B transition. 
In very thin slabs 0.3$\mu$m experiment \cite{Freeman:1988} did not observe the AB transition where it 
was expected from GL theory \cite{Li:1988tu} calculations. 
Later experiments \cite{Kawae:1998wx,Kawasaki:2004kp}
saw this transition and concluded that 
its approximate location is in agreement with calculations 
based on variational approach \cite{Fujita:1980jk}, 
but found that $^4$He coverage and change from diffuse to 
specular quasiparticle scattering moved the AB transition line to higher temperature, in 
disagreement with GL and QC theory. 
The most recent experiment in a high-precision single cell \cite{Levitin:2013hq} 
showed significantly lower $T_{AB}$ than predicted by weak-coupling quasiclassical 
calculations, but saw $T^{specular}_{AB} < T^{diffuse}_{AB}$ in agreement with 
quasiclassical theory prediction \cite{Nagato:2000vv}. 

Several other anomalies have been seen in transport experiments. 
Magnitude of the critical current in 
\cite{DAUNT:1988vn,Steel:1994dr} 
was significantly smaller than expected from early theoretical models, 
and flow rate transition was observed in \cite{Steel:1994dr}. 
Superfluid density in torsional oscillator \cite{XuCrooker} showed anomaly and some sort of transition 
when films became $D \lesssim 275$ nm. 
Third sound oscillation modes in \He\ films \cite{Schechter:1998ub} 
show unusual mode coupling. 
Film flow studies \cite{Saitoh:2003kz,Saitoh:2004tl,Saitoh:2007gp} 
reported two regimes of superflow dissipation in films with $D$ below and above $1\mu$m. 

These all call for deeper and wide-ranging investigation of states in films and slabs, 
to fully understand theoretical models and physical properties of phases in confinement.

\section{Conclusion}
\label{sec:last}

Unconventional superconductors and superfluids provide an 
excellent testing ground of our understanding of complex pairing condensates  
with multiple spontaneously broken symmetries. 
One of the special signatures of these states is extreme sensitivity 
of quasiparticle states to Andreev particle-hole scattering on interfaces 
that leads to formation of low-energy bound states concentrated in the region 
of several coherence length $\xi_0$ near interface. 

These states carry information about structure of the underlying 
order parameter and their properties, such as 
dispersion, weight, spin structure etc, encode the 
way these states were created: relative orientation of the interface with respect to 
the crystal lattice, scattering properties of a surface, or 
a particular way the incoming and outgoing trajectories connect the 
points of momentum-space Hamiltonian. 

Due to presence of midgap bound states and order parameter suppression, 
the surface region can have very different experimental signatures 
from those of the bulk phase. 
In confined geometry, influence of surface Andreev states is greatly 
enhanced because smallness of the volume precludes formation 
of dominant bulk phases. As a result, a new part of the condensate's phase space 
can be explored. 
Beside configurations that are trivially modified versions of the bulk states, 
several new phases are expected to be the ground states 
in confined geometry. The new phases have different 
symmetry properties, that depend on the nature of the 
pairing interaction. In single-component $d$-wave film spontaneous 
currents appear, breaking time-reversal symmetry. 
The currents are generated due to the fact that the bound states 
can lower their energy in the presence of a superflow. 
In superfluid \He\ multi-component nature of the condensate 
results in the amplitude modulation of the order parameter 
in the plane of the film, breaking continuous translation symmetry. 

Recent experimental progress in manufacturing of nanometer-scale cells 
promise a new window into extreme regimes of confinement,  
and a new approach to investigation of properties of 
surface bound states and unconventional condensates. 
Superfluid \He\ in confinement is a particularly interesting system 
where multiple phenomena challenge our understanding, 
both experimentally and theoretically. 
One question is whether or not additional phases may
be stabilized depending on the geometry and surface structure of
the confining geometry. So far there seems to be no experimental 
evidence for the proposed stripe phase in \He\ films. Detection of the 
non-uniform superconducting states, however, is a difficult task, as 
is evident from 50-year long search for Fulde-Ferrell-Larkin-Ovchinnikov 
phase in Pauli-limited superconductors. 
Another question, what is the origin of many anomalies seen in 
multiple experimental probes in confined \He? 
Is it simply a technical difficulty that is related to the complexity and 
extreme environment of the system, or there exist new uncovered physics that 
is related to the unusual quasiparticle states? 
We are at a point where one might have high expectations 
that the next few years will bring us answers to many of these questions.  

\vskip6pt




\ack{I would like to thank the Editors for inviting me to submit a manuscript for this 
theme issue. 
}

\funding{The work on this review is supported by NSF through grant DMR-0954342.}

\conflict{I have no competing interests.}



\bibliographystyle{rsta}
\bibliography{bibliography}

%
%
%

\end{document}